\documentclass[preprint,3p]{elsarticle}
\usepackage{epsfig}
\usepackage{amssymb}
\usepackage{graphicx}
\usepackage{epstopdf}
\usepackage{amssymb} 
\usepackage{dcolumn}
\usepackage{resize}
\usepackage{wrapfig}
\usepackage{picinpar}
\usepackage{enumitem}
\usepackage{multirow}
\usepackage{url}
\usepackage{bm}
\usepackage{amsmath}
\usepackage{relsize}
\usepackage[small,compact]{titlesec}
\usepackage[font=small]{caption}
\usepackage{subcaption}
\usepackage{amsthm}
\usepackage{tabularx}
\usepackage{color}
\def\beq{\begin{equation}}
\def\eeq{\end{equation}}

\def\etal{{\it et al.}}
\newcommand{\mylab}[3]{\raisebox{#2}[0mm][0mm]{%at
\makebox[0mm][l]{\hspace*{#1}#3}}}

\begin{document}
\begin{frontmatter}
\title{The effects of clinically-derived parametric data uncertainty in patient-specific coronary simulations with deformable walls}
\author[label1]{Jongmin Seo}
\ead{jongminseo@stanford.edu}
\author[label2]{Daniele E. Schiavazzi}
\author[label3]{Andrew M. Kahn}
\author[label1]{Alison L. Marsden}
\address[label1]{Department of Pediatrics (Cardiology), Bioengineering and ICME, Stanford University, Stanford, CA, USA}
\address[label2]{Department of Applied and Computational Mathematics and Statistics, University of Notre Dame, IN, USA}
\address[label3]{Department of Medicine, University of California San Diego, La Jolla, CA, USA}

\begin{abstract}
Cardiovascular simulations are increasingly used for non-invasive diagnosis of cardiovascular disease, to guide treatment decisions, and in the design of medical devices.
Quantitative assessment of the variability of simulation outputs due to input uncertainty is a key step toward further integration of cardiovascular simulations in the clinical workflow. 
In this study, we present uncertainty quantification in computational models of the coronary circulation to investigate the effect of uncertain parameters, including coronary pressure waveform, intramyocardial pressure, morphometry exponent, and the vascular wall Young's modulus. 
We employ a left coronary artery model with deformable vessel walls, simulated via an Arbitrary-Lagrangian-Eulerian framework for fluid-structure interaction, with a prescribed inlet pressure and open-loop lumped parameter network outlet boundary conditions. 
Stochastic modeling of the uncertain inputs is determined from intra-coronary catheterization data or gathered from the literature. 
Uncertainty propagation is performed using several approaches including Monte Carlo, Quasi Monte Carlo sampling, stochastic collocation, and multi-wavelet stochastic expansion. Variabilities in the quantities of interest, including branch pressure, flow, wall shear stress, and wall deformation are assessed.  
We find that uncertainty in inlet pressures and intramyocardial pressures significantly affect all resulting QoIs, while uncertainty in elastic modulus only affects the mechanical response of the vascular wall. 
% The morphometry exponent causes no change
Variability in the morphometry exponent used to distribute the total downstream vascular resistance to the single outlets, has little effect on coronary hemodynamics or wall mechanics. Finally, we compare convergence behaviors of statistics of QoIs using several uncertainty propagation methods on three model benchmark problems and the left coronary simulations. From the simulation results, we conclude that the multi-wavelet stochastic expansion shows superior accuracy and performance against Quasi Monte Carlo and stochastic collocation methods. 
\end{abstract}
\end{frontmatter}
%

% =====================================
\section{Introduction}\label{sec:intro}
% =====================================

% Cardiovascular models can help diagnostics
\noindent Cardiovascular disease is the leading cause of death worldwide with projected total cost of over 1 trillion dollars by 2035, according to the American Heart Association~\cite{AHA2018}.
Cardiovascular modeling provides non-invasive tools to complement clinical diagnostics, and patient risk assessment as well as predictive capabilities to aid in clinical decision-making and treatment planning.
Modern analysis and simulation of blood flow in the cardiovascular system requires a combination of clinical data acquisition, image processing for anatomic model construction, selection of appropriate physiologic boundary condition and wall material properties, accurate solution of the governing equations, and high-performance computing~\cite{Marsden2014, Marsden2015}.
Clinical applications of cardiovascular simulation include the non-invasive assessment of fractional flow reserve FFR$_\text{CT}$~\cite{Taylor2013}, which is highly correlated with an invasive measurement of FFR and has drawn a great deal of attention as a reliable predictor of obstructive coronary disease and ischemia~\cite{Baumann2015, KOO20111, Min2012, Douglas2015, Kurata2017}.
Moreover, cardiovascular models have been used for new surgical designs for congenital heart disease~\cite{yang2010constrained}, risk assessment in coronary artery bypass surgery~\cite{Ramachandra2017}, thrombotic risk stratification in Kawasaki disease~\cite{Noelia2017}, ventricular assist device design~\cite{Long2014}, aneurysm treatment~\cite{Frauenfelder2006}, and stent design~\cite{migliavacca2002mechanical, Gundert2012}.

However, current cardiovascular models typically provide only deterministic predictions while ignoring variability of simulation outputs due to numerous uncertainties in cardiac catheterization data, tissue properties, medical imaging, and boundary condition selection. 
In this context, as simulation data are increasingly incorporated into cardiovascular disease research, clinical trials, and the FDA approval process, there is a pressing need to establish strict guidelines and effective methods to assess the impact of uncertainty on simulation predictions. 
%
%A patient-specific cardiovascular modeling process involves multiple tasks each of which is prone to involve uncertainty. The effect of uncertainty is typically left undiscussed when reporting results.
%
Cardiovascular models are increasingly assimilating data from a range of clinical imaging and data sources, obtained both invasively and non-invasively.    
Uncertainty is present in common non-invasive clinical data measurements (e.g., heart rate or blood pressure), echocardiography (e.g., stroke volume, ejection fraction, cardiac output, acceleration times), image data (CT or MRI). It is also present in invasively obtained cardiac catheterization data (pressures, intravascular ultrasound).   
Additional uncertainties stem from the vessel-wall histology and material properties. 
% Same patient uncertainty and population variability
Inter-patient variability reported in population studies as well as intra-patient variability in repeated measurements (when available), can be used to inform the amount of uncertainty in the simulation inputs or modeling parameters. 

Once variabilities in the input parameters are assessed, uncertainty quantification (UQ) tools provide a means to investigate the relationship between the input and output variabilities. 
Reporting mean and confidence intervals as well as higher statistical moments for clinically relevant quantities is key for translation of simulation tools to the clinic.
However, the computational cost of UQ is a major bottleneck, typically requiring solution of an expensive model at each realization of a potentially large collection of random inputs.
This cost may further increase with deformable models, including fluid-structure interaction~\cite{Figueroa2006} or with the inclusion of physiologic boundary conditions~\cite{Esmaily2013res} possibly assimilated from available clinical data under uncertainty~\cite{Tran2017, Schiavazzi2017}.
In this context, several one-dimensional cardiovascular models, assuming blood as a Newtonian fluid and fully-developed, axisymmetric flow inside a cylindrical vessel have been employed previously to demonstrate successful integration of UQ in cardiovascular modeling.
Taking advantage of the low computational cost of one-dimensional hemodynamic models, several studies discussed the effect of variability in constitutive model parameters~\cite{Xiu2007}, arterial wall stiffness, inlet velocity~\cite{Brault2016}, geometry, resistance and pressure~\cite{Chen2013}, and assessment of global sensitivity.
One-dimensional models are however limited when one wishes to understand and quantify realistic flow patterns and local flow features.  

Sankaran and Marsden~\cite{Sankaran2010} proposed one of the first studies in uncertainty quantification applied to three-dimensional cardiovascular models, focusing on the effect of uncertainties in the shape optimization of a bypass grafts geometry. 
In addition, Sankaran and Marsden~\cite{Sankaran2011} applied UQ on both idealized and patient-specific models of Fontan surgical configurations including clinically relevant parameter uncertainty, such as the geometry, the inlet velocity, and the flow-split between the left and right pulmonary arteries.
While these early studies assumed {\it a priori} distributions for the uncertain parameters, Schiavazzi \etal~\cite{Schiavazzi2016} leveraged clinical measurements of flow split and pulmonary pressure to assimilate distributions of outlet resistances from the solution of a Bayesian inverse problem in model configurations related to specific stages of single ventricle palliation surgery.
More recently, Sankaran \etal~\cite{Sankaran2016} focused on the effect of multiple sources of uncertainty on the non-invasive estimation of FFR$_{\text{CT}}$ from cardiovascular models, and identified the minimum lumen diameter uncertainty as a major contributor to the FFR$_{\text{CT}}$ variability. 

In the majority of the above-mentioned studies, the propagation of uncertainty through the cardiovascular models was performed using the Stochastic Collocation method (SC), which was often regarded as an accurate choice, particularly for smooth parametric response surfaces. 
However, the smoothness of the stochastic response is not known \emph{a priori}, therefore it is preferred to use a UQ approach that works equally well under smooth and non-smooth conditions. 
In this context, Schiavazzi \etal~\cite{Schiavazzi2014,Schiavazzi2016} proposed a generalized multi-resolution stochastic expansion framework which performs successive binary refinements in the stochastic domain, builds multi-wavelet families orthogonal to the PDF of the random inputs in each subdivided element of the parameter space, and leverages relevance vector machine regression~\cite{Tipping2001} to minimize the number of model evaluations needed to determine accurate statistics of the quantities of interest.
The MW approach has been shown to perform better than MC, QMC and SC in various benchmarks, and applied to coronary artery bypass graft modeling~\cite{Schiavazzi2016, Tran2019}.

Many studies in the literature have focused on rigid wall models, due to the large computational cost of deformable wall models and the complexity of accounting for the interaction between fluid and structure. However, wall deformation is critical for capturing vascular wave propagation, and for characterizing mechanical forces acting on the vessel wall such as wall shear stress which are crucial to mechanobiology. In coronary arteries, wall strain is considered to be a key quantity affecting the progression of atherosclerosis \cite{Gijsen2013}.     
In most cases, arterial wall stiffness is unknown {\it in vivo}, spatially inhomogeneous (e.g. different for femoral, carotid, and coronary arteries~\cite{Gow1979}, and even within in the same anatomical region), and varies with the presence and severity of atherosclerosis~\cite{Hirai1989, Karimi2013}.
Additionally, material properties are affected by significant uncertainty with only limited available data, as indicated by measurement errors from postmortem vascular stiffness~\cite{Gow1979}, and intra-/inter-patient variability~\cite{Gow1979, Karimi2013}. 
The effect of material property uncertainty on one-dimensional hemodynamics was considered in~\cite{Xiu2007, Chen2013, Brault2016}, and Biehler \etal~\cite{Biehler2015, Biehler2017} studied the effects of such uncertainties in the biomechanics of an aortic abdominal aneurysm using a multi-fidelity approach. 
To date, Tran \etal~\cite{Tran2019} was among the first studies to consider material property uncertainty in a three-dimensional patient-specific model incorporating fluid-structure interaction.
In this work, fluid-structure interaction was enabled using a lightweight shell formulation via the coupled-momentum method (CMM)~\cite{Figueroa2006}. 

This study presents an uncertainty quantification based on a three-dimensional patient-specific cardiovascular model with Arbitrary Lagrangian-Eulerian (ALE) fluid-structure interaction.
ALE is a realistic and computationally intensive approach to solve the Navier-Stokes equations, the equations for solid mechanics, and the mesh motion equations in a monolithic fashion. 
We take advantage of recent developments in the SimVascular open-source software package~\cite{Lan2018}, namely the svFSI flow solver which provides an implementation of ALE FSI.
The aim of the study is threefold. 
First, we aim to demonstrate the application of various UQ propagation strategies to a realistic patient specific model with ALE. To make uncertainty propagation feasible, we exploit sub-modeling strategies to reduce the prohibitive cost of UQ in realistic models. In addition, we used the best performing linear solvers among other available solvers tested in our previous study, to reduce the computational burden in uncertainty propagation. 
Second, we quantify how uncertainty is amplified or reduced from the inputs to the outputs of the selected model. We consider several sources of uncertainty including pulsatile coronary pressure waveforms, intramyocardial pressure, vessel wall elasticity, and morphometry exponent. 
These quantities are highly relevant in coronary artery disease simulation, and therefore of key importance for validation and verification. 
In this context, it has been previously shown that FFR and wall shear stress in coronary arteries may be subject to significant variability as a result of perturbations in inlet and outlet boundary conditions~\cite{Kurata2017, Boileau2018, GIESSEN2011}.
Third, we assess and compare the performance of multiple uncertainty propagation algorithms specifically applied to deformable coronary sub-models.

The paper is organized as follows. In Section~\ref{sec:LCAmodeling}, we discuss governing equations, generation and application of boundary conditions to our deformable left coronary artery model.
In Section~\ref{sec:UQmodeling}, we introduce uncertainties in coronary simulation parameters, and review both the formulation and properties of several approaches for uncertainty propagation from the literature.
The uncertainty propagation results are discussed in Section~\ref{sec:results}.
We first focus on the results of various UQ techniques to three benchmark problems with closed-form analytic formulation, and proceed to show how uncertainty in the selected coronary simulation parameters affect a number of relevant outputs, including pressure, flow, wall shear stress and measures of mechanical response at the wall. 
We conclude with an additional discussion in Section~\ref{sec:Discussion}, that includes study limitations and future work.

% ========================================================================
\section{Sub-models for the coronary circulation with deformable walls}\label{sec:LCAmodeling}
% ========================================================================

\begin{figure}[ht]
\vspace{-6pt}
\centering
\includegraphics[width=0.9\textwidth, keepaspectratio]{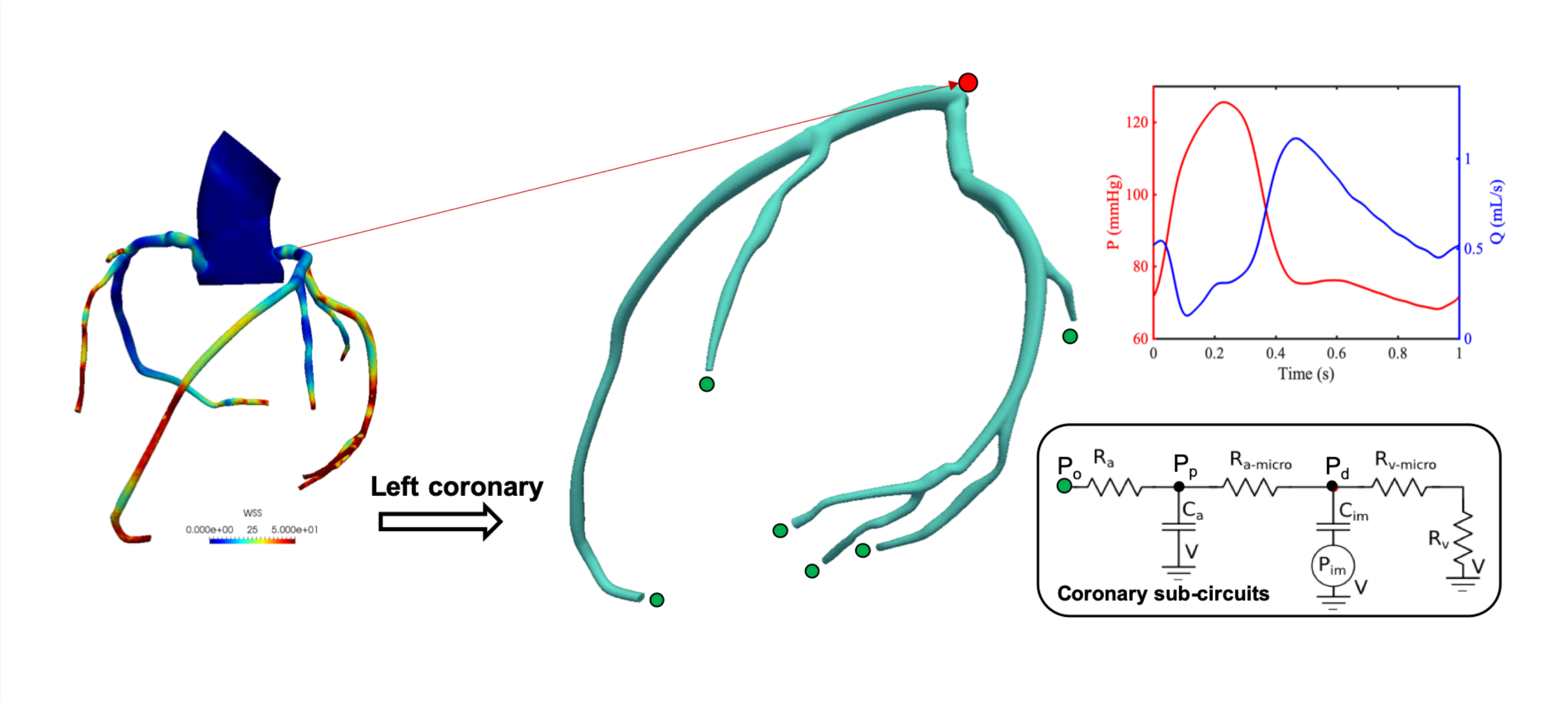}
\mylab{-144mm}{49mm}{(a)}%
\mylab{-46mm}{58mm}{(b)}%
\mylab{-50mm}{28mm}{(c)}%
\caption{(a) Left coronary artery (LCA) sub-model and geometrically multi-scale patient-specific aorto-coronary model.
(b) Typical coronary pressure and flow outputs from LCA model. 
(c) Lumped parameter boundary conditions prescribed at each outlet.}\label{fig:1}
\end{figure}

\noindent In this study, we investigate the response of a three-dimensional aorto-coronary model, representing right and left coronary arteries and including a segment of the aortic arch shown in Figure~\ref{fig:1}(a). 
The anatomy was constructed from CT clinical images using SimVascular~\cite{Updergrove2016}. The aorto-coronary model has lumped-parameter network (LPN) boundary conditions at the coronary artery branches and outlet of the aorta. At the inlet of the aorta, a physiologic pulsatile flow waveform is prescribed. 
The LPN parameters are tuned to match multiple clinical targets (stroke volume, ejection fraction, pressure etc.) and confirmed to reproduce physiologically admissible responses~\cite{Tran2017}. 

To minimize computational effort while still providing accurate coronary response, our study focuses on a left coronary sub-model, where appropriate boundary conditions are applied to match the simulation outcomes of the complete model (e.g., a similar approach in~\cite{Tran2019}).
This reduces the total volume of the region of interest to approximately 1/15 of the full aorto-coronary model.
The pressure waveform P$_\text{in}$ at the inlet of the submodel was extracted from the full model simulation results, while all downstream boundary conditions were kept the same. 
The lumen diameter of the left main coronary is 4 mm and the diameter of the distal left anterior descending artery (LAD) is 2.7 mm, consistent with the range of diameters on normal human left coronaries, the proximal LAD diameter d=3.7$\pm$0.4 mm, and distal LAD d=1.9$\pm$0.4 mm~\cite{Perry2013, Dodge1992}. 

\begin{figure}[t]
\vspace{-6pt}
\centering
\includegraphics[width=0.9\textwidth, keepaspectratio]{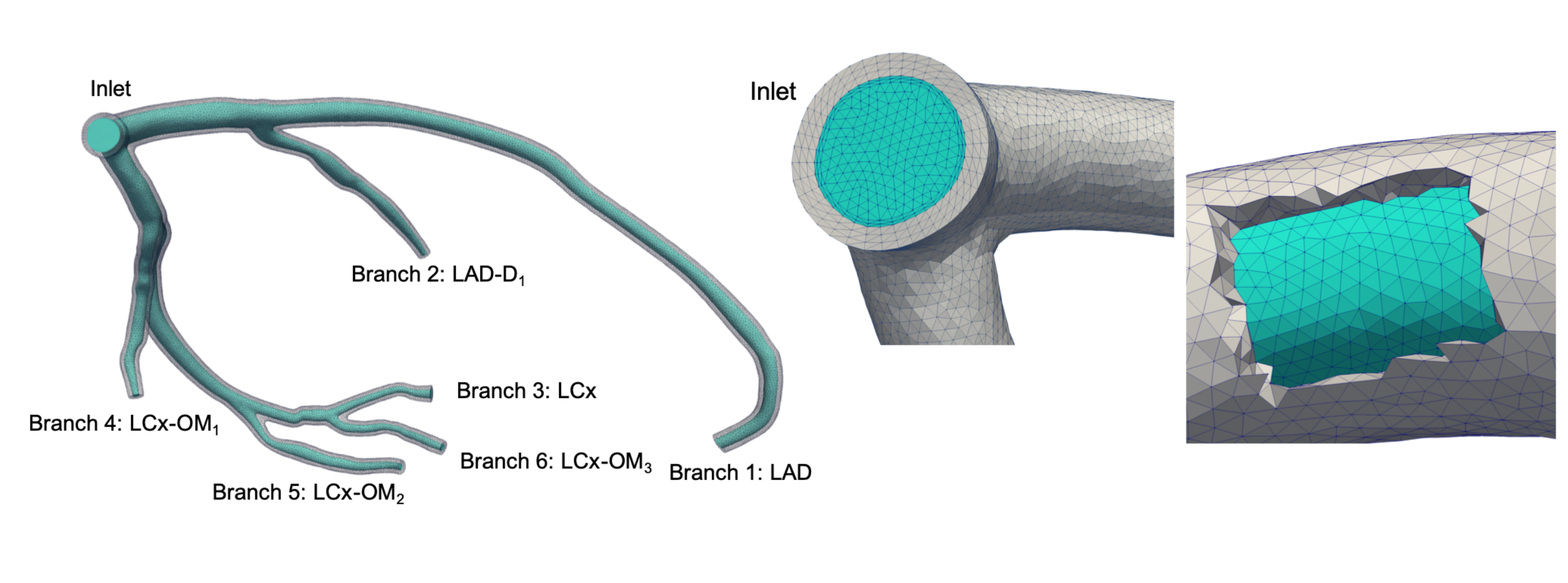}
\mylab{-147mm}{44mm}{(a)}%
\mylab{-85mm}{44mm}{(b)}%
\mylab{-36mm}{41mm}{(c)}%
\caption{LCA computational mesh with deformable walls. The lumen mesh is colored with cyan and the wall mesh is colored with gray. 
(a) LCA model with branch naming conventions (b) inlet mesh and (c) a cut-through view of the vessel wall mesh.}\label{fig:2}
\end{figure}

The wall structural mesh was obtained using a uniform wall thickness h=0.08 mm (Figure~\ref{fig:2}). 
The coronary wall thickness is consistent with two electrocardiographic studies that consistently reported a coronary wall thickness of 1.0$\pm$0.2 mm~\cite{Perry2013, Pizlo2003} and with the morphometric law for coronary vessel thickness by Podesser~\cite{Podesser1998}, and larger than a typical vessel wall thickness equal to 10\% of the vessel radius found in the literature~\cite{McDonald2011, Bergel1960, Pearson1994, Peterson1960, Patel1969}.

Interaction between fluid and structure is simulated through Arbitrary Lagrangian-Eulerian (ALE) coupling provided by the SimVascular svFSI solver, which implements a variational multi-scale finite element method with second order implicit generalized-$\alpha$ time integration~\cite{Bazilevs2007, Esmaily2015}. The incompressible Navier-Stokes equations in ALE form are
\begin{equation}
\begin{split}
\rho\,\frac{\partial \mathbf{u}}{\partial t}|_{\hat{ \mathbf{x}}}+\rho\,{\mathbf{v}}\cdot\nabla\,{\mathbf{u}}  &= \rho\,\mathbf{f}+ \nabla \cdot {\mathbf{\sigma}_f}\\
\nabla  \cdot {\mathbf{u}} &= 0
\end{split}
\quad\text{in}\,\,\Omega_f,
\end{equation}
where $\rho, \mathbf{u}=\mathbf{u}(\mathbf{x},t)$, and $\mathbf{f}$ are fluid density, velocity vector, and body force, respectively, in the fluid domain $\Omega_f$. \
We model blood as a Newtonian fluid, for which $\mathbf{\sigma}_f=-p\,{\mathbf{I}} + \mu\,(\nabla{\mathbf{u}}+\nabla{\mathbf{u}}^T) = -p\,{\mathbf{I}} + \mu\,\nabla^{s}{\mathbf{u}}$, with $\mu$ the kinematic viscosity, $p=p(\mathbf{x},t)$ the pressure, and $\mathbf{v}=\mathbf{u-\hat{u}}$ the fluid velocity relative to the mesh.
Additionally, in the solid domain $\Omega_s$, we solve the equilibrium equation, 
\begin{equation}
\begin{split}
\rho_s\,{\frac{\partial \mathbf{u}}{\partial t}}=\rho_s\,\mathbf{f} + \nabla\cdot \sigma_s
\end{split}
\quad\text{in}\,\,\Omega_s,
\end{equation}
where $\rho_s$ and $\sigma_s$ denote the density and solid stress tensor, respectively. 
The spatial discretization is based on the variational multi-scale method \cite{Bazilevs2007, Esmaily2015, Seo2019}. We use the Saint Venant-Kirchhoff hyper-elastic constitutive model. We employ P1-P1 (linear and continuous) finite elements for velocity and pressure. {The weak formulation of the above equations and process leading to an algebraic system of equations has been diascussed in~\cite{Esmaily2015, Seo2019}}.
% Linear solver
The solution of the linear algebraic system is computed using the Trilinos library~\cite{Trilinos}, developed at Sandia National Laboratory and coupled with the SimVascular svFSI solver.
We use either the Bi-Conjugate Gradient iterative linear solver with incomplete LU preconditioner or the Generalized Minimum Residual with a diagonal preconditioner. These combinations were shown to be optimal for cardiovascular simulations with deformable walls~\cite{Seo2019}.

Coronary boundary conditions~\cite{Kim2009, Sankaran2012} are enforced at the $n_{o}=6$ left coronary artery (LCA) outlets, designed to capture the diastolic nature of the coronary flow. These consist of a LPN circuit governed by the ordinary differential equations
\begin{equation}
\frac{dP_\text{p,i}}{dt}=\frac{1}{C_\text{a,i}}\Big(Q_{i}-\frac{P_\text{p,i}-P_\text{d,i}}{R_\text{am,i}}\Big),\;\;i=1,2,...,n_o. 
\end{equation}
\begin{equation}
\frac{dP_\text{d,i}}{dt}=\frac{1}{C_\text{im,i}}\Big(\frac{P_\text{p,i}-P_\text{d,i}}{R_\text{am,i}}-\frac{P_\text{d,i}-P_\text{im}}{R_\text{v,i}}\Big)+\frac{dP_\text{im}}{dt}, \;\;i=1,2,...,n_o. 
\end{equation}
where $P_\text{p,i}$ and $P_\text{d,i}$ are the proximal and distal pressures, $C_\text{a,i}$ and $C_\text{im,i}$ are the proximal and distal capacitances and $R_\text{am,i}$ and $R_\text{rv,i}$ are the resistances, respectively, as shown in Figure~\ref{fig:1}(c). 
Note that the same rate of intramyocardial pressure $dP_\text{im}/dt$ is prescribed equally at all the $n_o$ outlets. 
The above equations are integrated in time using the fourth-order Runge-Kutta method at each time step, with the final outlet pressure $P_\text{o,i}$ computed as $P_\text{o,i}=P_\text{p,i}+R_\text{a,i}\,Q_\text{i}$ and coupled to the three-dimensional model solution~\cite{Esmaily2012}. 

The total coronary resistance was computed assigning 4$\%$ of the cardiac output to the coronary arteries~\cite{Bogren1989} and was distributed among outlets following a morphometric relation associating flow rates with vessel diameters, i.e., $Q\propto (d/2)^{m}$. The diameter is approximated using the square root of the area $\sqrt{A_{i}}$.
The resistance of a distal branch, $R_{i}$, is therefore computed as 
\begin{equation}
R_i=\frac{\sum_{j} \sqrt{A_j^m}}{\sqrt{A_i^m}}\cdot R_\text{total},\;\;\;i,j=1,2,...,n_o,
\label{eq:R}
\end{equation}
while the capacitances are instead distributed proportional to the outlet area~\cite{Sankaran2012}.

Fully fixed mechanical restraint conditions were applied at the inlet and outlets, while a zero Neumann pressure was applied on the outer vascular wall. 
In addition, the diastolic configuration, which is acquired from the CT scans and used to reconstruct the model geometry, is assumed stress-free, no pre-stress analysis was performed \cite{Hsu2011}, and therefore no uncertainty has been quantified for this pre-stress.
This is consistent with our selection of the displacement as the preferred mechanical quantity of interest. We note that pre-stress has been considered in recent work \cite{Hsu2011, Baumler2020} and could be a target of uncertainty analysis in future studies.

In the sub-modeling process, we first extracted an area-averaged flow waveform resulting from the full model at the location of the submodel inlet, and applied it to the submodel inlet as a Dirichlet boundary condition. We then verified the excellent agreement between the flow and pressure distributions resulting for the two models.
After verifying the submodel for a prescribed inlet flow, we switched to a pulsatile pressure boundary condition at the inlet so that a change in the intramyocardial pressure at the outlets could affect the flow through the LCA. 

% Computational Mesh
Figure~\ref{fig:2} shows the computational mesh for this study, with 567,373 tetrahedral elements for the left coronary artery lumen.
% Convergence study
This optimal mesh size resulted from a convergence study where we compared discretizations with $\approx$ 500K, 1M, 2M, and 4M isotropic tetrahedral meshes, and $\approx$ 500K, 1M, and 2M meshes with boundary layers. 
We find that none of the outlet pressures and flow rates vary by more than 3 percent, across all tested meshes except for the 500K isotropic  mesh. 
Time-averaged wall shear stress (TAWSS) was significantly affected by the mesh resolution near the wall. 
With a coarse wall mesh size with isotropic meshing, TAWSS is underestimated even with a large total number of grid points (e.g. 4M).
Our 500K boundary layer mesh provided instead the correct TAWSS within 3 percent root-mean-square error against the finest model with boundary layers. The wall mesh contains 373,435 tetrahedral elements, with three elements through the thickness, which is regarded as appropriate due to the prevalent membrane deformations (through thickness bending is assumed to be small).
We employed the Meshmixer application (\texttt{http://www.meshmixer.com/}) to enforce the nodal continuity at the fluid-solid mesh interface, as shown in Figure~\ref{fig:2}(c)~\cite{Seo2019, Vedula2017, Baumler2020}. 

% ============================================
\section{Uncertainty modeling and propagation methods}\label{sec:UQmodeling}
% ============================================

\noindent In this section, we define several sources of uncertainty relating to quantities that are essential for physiological admissibility of the simulation results.
Uncertain inputs are treated as random variables or random vectors, with assumed underlying distributions. 
We chose to focus on the coronary artery inlet pressure waveform, the intramyocardial pressure, the morphometry exponent used in the assignment of outlet boundary conditions, and the material properties.
Uncertainty in the coronary artery pressure is determined from clinical data acquired \emph{in-vivo} through cardiac catheterization, while uncertainty in the morphometry and the material property are determined from a thorough literature review.

% =======================================================================
\subsection{Uncertainties in coronary artery pressure}\label{sec:coronaryPresUncertainty}
% =======================================================================

\begin{figure}[t]
\vspace{-6pt}
\centering
\includegraphics[width=1.0\textwidth, keepaspectratio]{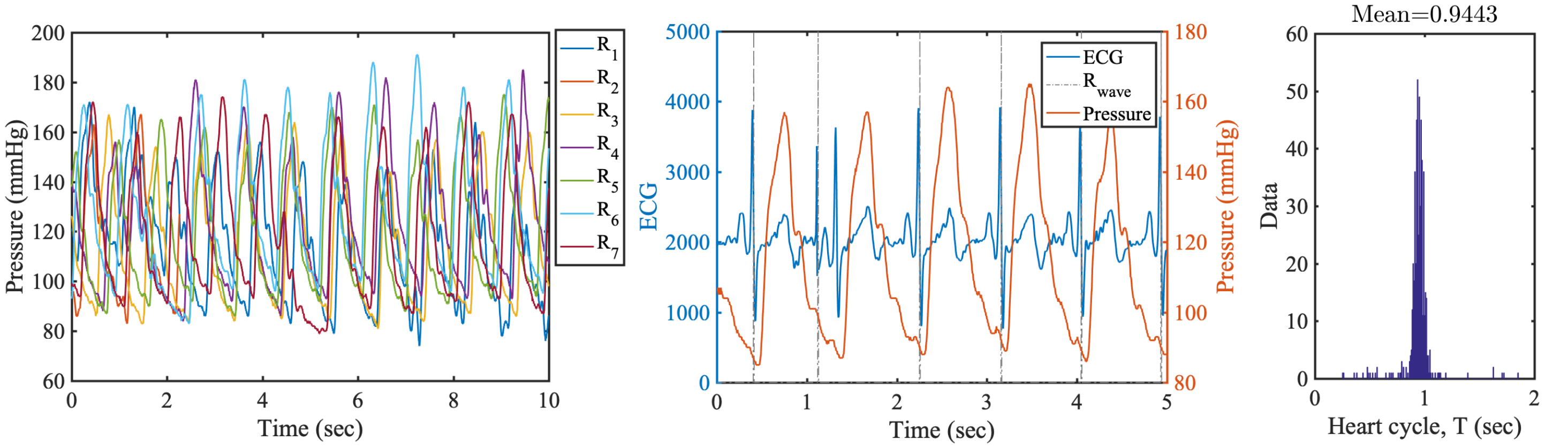}
\mylab{-84mm}{50mm}{(a)}%
\mylab{-17mm}{50mm}{(b)}%
\mylab{48mm}{50mm}{(c)}%
\caption{Clinical data from cardiac catheterization. (a) Seven measurements of pulsatile pressure waveforms in the coronary artery from Patient 1. The labels R$_1$-R$_7$ represent different measurements. (b) Pulsatile pressure waveforms overlaid with the patient ECG. A starting instance of the cardiac cycle is recorded by R$_\text{wave}$. (c) A cardiac cycle duration histogram for Patient 1.}\label{fig:3}
\end{figure}

\noindent Uncertainty stems from cardiac catheterization in pressure measured by inter- and intra-patient physiologic variability, as well as systematic measurement errors. We collected intra-coronary pressure data from six patients. Intra-coronary pressure waveforms and ECG signals were acquired during cardiac catheterization performed at the UCSD Medical Center, using a ComboWire (Volcano Inc.) catheter over more than 100 cardiac cycles. Patient characteristics are shown in Table ~\ref{table:1}. An example of pressure time-trace with overlaid ECG signal is shown in Figure \ref{fig:3}. 
\begin{table}[ht]
\centering
\begin{tabular}{c | c c c c c c c c c }
\hline
Patient & Sex & Age & Weight(kg) & Height(cm) & BMI & Location & Lesion \\
\hline
1&Male&74&127&170&45.3&LAD&Proximal 50$\%$ stenosis \\
2&Male&68&88&188&24.9&LAD&None\\ 
3&Male&77&91&170&31.5&LAD,LCx& None\\
4&Male&70&47&155&19.7&LAD,LAD,RCA& None\\
5&Female&52&98&185&28.5&RCA& Ostial 50$\%$ stenosis\\
6&Male&49&67&170&23.2&RCA& Mid 50$\%$stenosis\\
\hline
\end{tabular}
\caption{Characterization of the patient cohort included in the present study. LAD: left anterior descending artery; LCx: left circumflex artery; RCA: right coronary artery.}
\label{table:1}
\end{table}

\begin{figure}[ht]
\vspace{-6pt}
\centering
\includegraphics[width=0.9\textwidth, keepaspectratio]{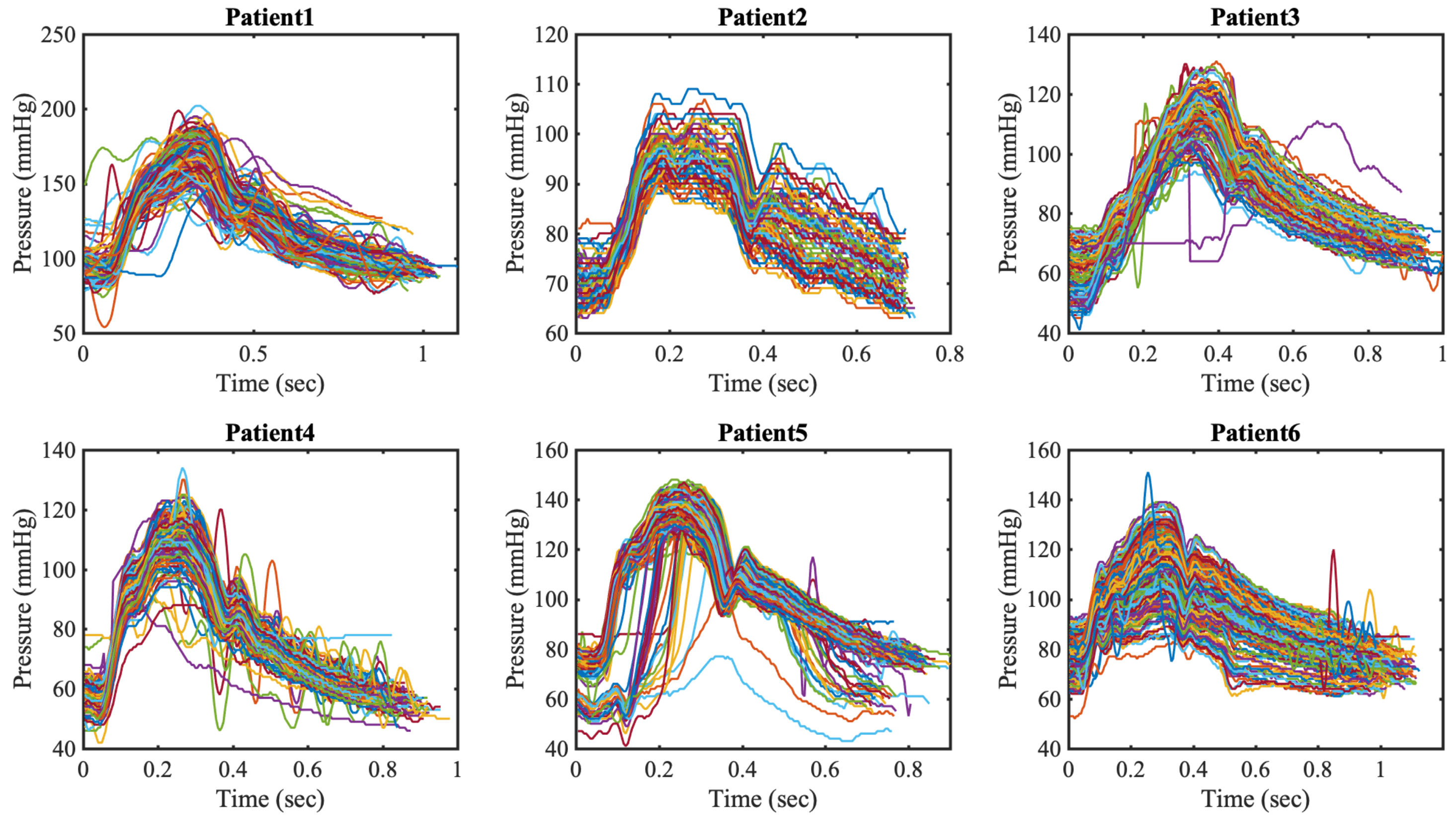}
\caption{Post-processed intra-coronary pressure waveforms from six patients. Each pressure waveform is plotted in the associated heart cycle.}\label{fig:4}
\end{figure}

We registered the pressure waveform to lie within the cardiac cycle as shown in Figure~\ref{fig:3}(b). The starting instance R$_\text{wave}$ was recorded and used to calculate the heart cycle. 
The histogram of heart cycle durations for Patient 1 in Figure~\ref{fig:3}(c) shows a narrow distribution around the sample mean value, i.e., T$\approx 0.94$, with a small number of outliers associated with durations T$\lesssim0.8$ or T$\gtrsim1.2$. By tracking individual durations, we confirmed that the outliers were generated from measurement errors rather than physiologic variability.
Thus, we rejected data with duration less than 80 percent and larger than 120 percent of the sample mean, capturing most data without errors. With the method described above, we post-processed pressure waveforms as shown in Figure~\ref{fig:4}.

\begin{figure}[ht!]
\vspace{-6pt}
\centering
\includegraphics[width=0.95\textwidth, keepaspectratio]{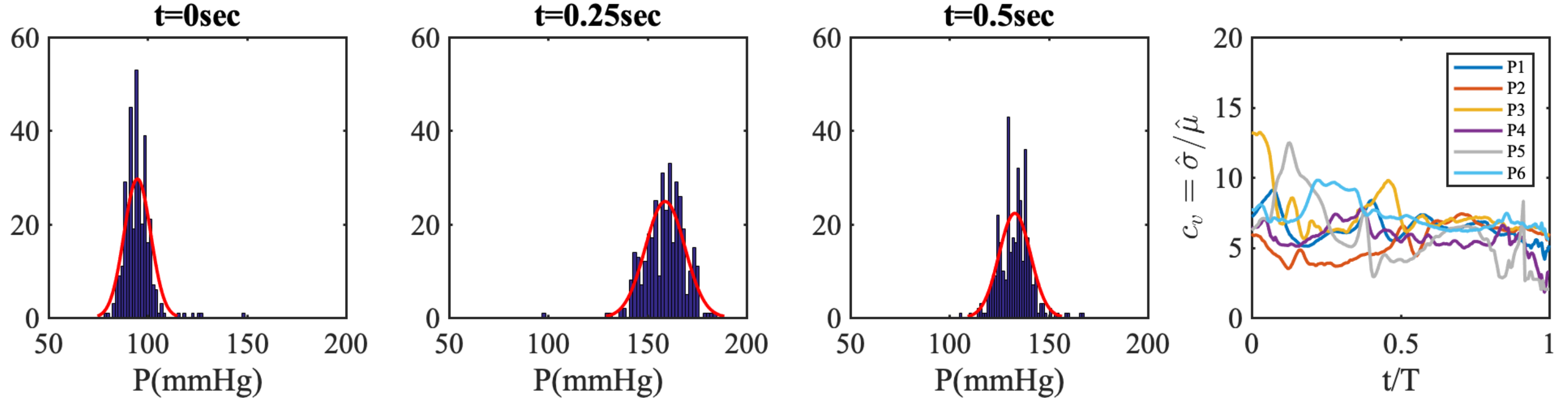}
\mylab{-161mm}{38mm}{(a)}%
\mylab{-120mm}{38mm}{(b)}%
\mylab{-81mm}{38mm}{(c)}%
\mylab{-41mm}{38mm}{(d)}%
\caption{(a-c) Histograms of Patient 1 pressure data at four points in time. The red lines represent a Gaussian distribution fit using the first two sample moments. (d) The time-history of coefficient of variation, $c_v$ for patients 1 to 6.}\label{fig:5}
\end{figure}

From the intra-coronary pressure data at each point in time, we first took ensemble average of all pressure values and denoted ensemble averaged mean as $\hat{\mu}$, standard deviation as $\hat{\sigma}$, and the coefficient of variation as $c_v={{\hat{\sigma}}/{\hat{\mu}}}$. Then, we averaged the ensemble averaged quantities over time and denoted the time-averaged quantities with bar superscript. While intra- and inter- patient physiologic variability was clearly observed in the mean and extreme pressures for each patient, we note that the time average of coefficient of variation, $\overline{c}_{v}$, shows a similar range equal to 5 to 7 percent across all patients (Table~\ref{table:2}).
To better identify the distribution characterizing the variability of coronary artery pressure at a single point in time, we plotted histograms at three discrete time instances in Figure~\ref{fig:5}(a-c), observing a symmetric and bell-shaped data distribution well fitted by a Gaussian. 
In addition, Figure~\ref{fig:5}(d) shows the time-history of the coefficient of variation over one heart cycle for all patients included in this study. For most patients, the coefficient of variation was found to be well approximated by a constant. As a result we employed a non-stationary Gaussian model with time-dependent mean equal to average measured pressure and time-independent variance.

\begin{table}[ht!]
\centering
\begin{tabular}{c  c c c c c c c c c }
\hline
Patient&& \multicolumn{6}{ c }{Pressure (mmHg)} && Heart cycle (sec) \\
 \cline{3-8}  \cline{10-10} 
No.&&Max.&Min.&&Mean&Std.&$\overline{c}_{v}$&&Mean\\
 \cline{1-1} \cline{3-4} \cline{6-8} \cline{10-10}
1&&202&74&&122&7.8&6.4&&0.95\\
2&&109&63&&82&4.4&5.3&&0.70\\
3&&131&41&&83&6.2&7.4&&0.91\\
4&&134&42&&76&4.4&5.8&&0.83\\
5&&148&41&&98&6.0&6.1&&0.79\\
6&&139&52&&92&6.8&7.4&&0.94\\
\hline
\end{tabular}
\caption{Intra-coronary pressure data measured from a cardiac catetherization on six patients. Pressures are in mmHg, heart cycles are in second, and $\overline{c}_v$ is reported as a percentage.}
\label{table:2}
\end{table}

\begin{figure}[t]
\vspace{-6pt}
\centering
\includegraphics[width=1.0\textwidth, keepaspectratio]{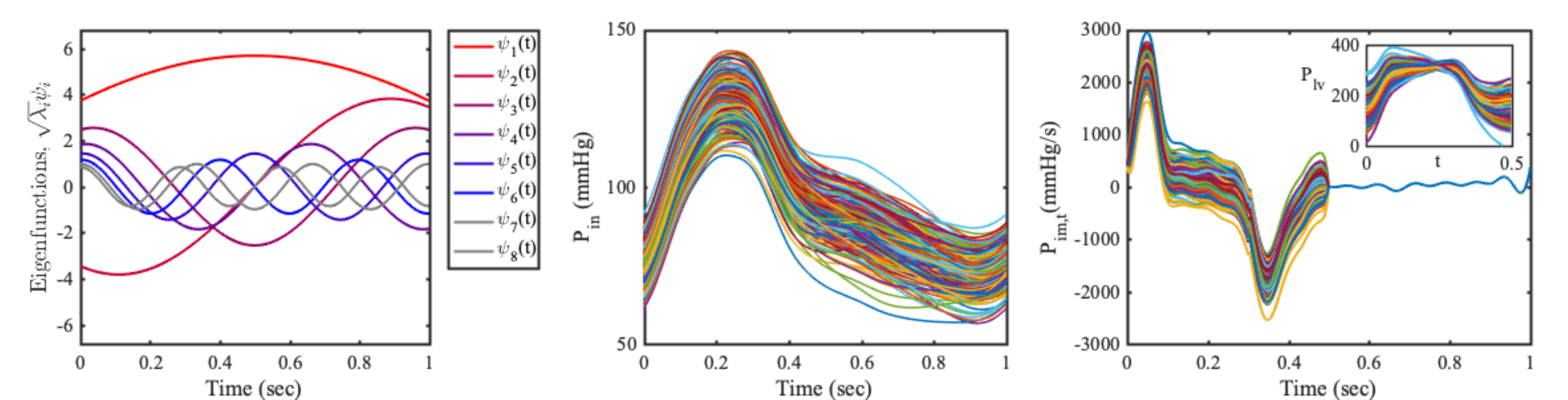}
\mylab{-80mm}{44mm}{(a)}%
\mylab{-23mm}{44mm}{(b)}%
\mylab{27mm}{44mm}{(c)}%
\caption{Stochastic modeling using the Karhunen-Lo\`eve expansion. (a) Eight modes of scaled eigenfunctions from the exponential covariance function, $\text{\bf K}(t,t')$. (b) 100 realizations of the perturbed coronary pressure waveform P$_\text{in}$, (c) 100 realizations of the perturbed time derivative of the intramyocardial pressure,  P$_\text{im,t}$(t). Inset: Reconstructed left ventricular pressure waveforms.}\label{fig:6}
\end{figure}

For our forward propagation study, we use a Karhunen-Lo\`eve (K-L) expansion for the coronary pressure waveform for each patient, resulting in a Gaussian process parameterized through multiple independent Gaussian random variables~\cite{ghanem2003stochastic}.
Similar approaches have been previously demonstrated in the literature. For example, a spectral decomposition is proposed in~\cite{Brault2016} to model the random distribution of aortic stiffness in a one-dimensional cardiovascular model, while the Expansion Optimal Linear Estimation (EOLE) approach is used in~\cite{Tran2019} to model the spatial distribution of material properties in a coronary bypass graft.
In this study, we assume an exponential covariance function, $\text{\bf K}(t,t')=\sigma^2 \text{exp}(-|t-t'| / l_c)$, where $t$ and $t'$ are two arbitrary time points, $l_c$ is the correlation length, and $\sigma^2$ the process variance. 
The Karhunen-Lo\`eve expansion of the stochastic process $\text{P}(t,\omega)$ is
\begin{equation}\label{eq:2}
\text{P}(t,\omega)=\hat{\text{P}}(t)+\sum_{i=1}^N \sqrt{\lambda_i}\,\psi_i(t)\,\xi_{i}(\omega),
\end{equation}
where $\boldsymbol{\xi}(\omega) = (\xi_1(\omega),\xi_2(\omega),\dots,\xi_N(\omega))$ is a collection of independent standard Gaussian random variables, $\lambda_i$ and $\psi_i$ are the eigenvalues and eigenvectors of the selected covariance kernel, $\text{\bf K}(t,t')=\sum^N_{i=1} \lambda_i \psi_i(t) \psi_i(t')$, respectively \cite{Maitre2002}, and $N$ is the truncation level, the number of modes included in the expansion.
After examining the covariances for the six patients, we selected a correlation length $l_c=\text{T}/2=0.5$ sec, which produced a satisfactory approximation of $\text{P}(t,\omega)$ using only the first four eigenmodes.
Specifically, the largest scaled eigenvalue spectra of the covariance function, $\lambda_i/\max_{i}(|\lambda_i|)$, are $\lambda_2/\lambda_1=34\%$, $\lambda_3/\lambda_1=14\%$, $\lambda_4/\lambda_1=7\%$, and decays to less than 5 percent after the first four modes. 
We set the process variance to 7 percent of the mean value as observed from the cardiac catheterization data. 
Samples from $\xi_{i}(\omega)$, $i=1,\dots,4$ are obtained by projecting a four-dimensional Sobol sequence through the inverse multivariate Gaussian cumulative distribution function. 
We plot a finite collection of eigenfunctions from the covariance in Figure~\ref{fig:6}(a), and realizations from the identified pressure stochastic process in Figure~\ref{fig:6}(b).

% =========================================================================
\subsection{Uncertainty in the intramyocardial pressure}\label{sec:intramyoPresUncertainty}
% =========================================================================

\noindent In cardiovascular modeling, specialized boundary conditions are typically employed to model coronary physiology~\cite{Kim2009}. 
While the contraction of the cardiac muscle impedes coronary flow during systole, coronary arterial flow rates increase and reaches their maximum during diastole following relaxation of the heart muscle.
The coronary flow waveform is therefore out-of-phase with the aortic pressure which is maximum at systole (see Figure \ref{fig:1}(b)).
To model this effect, a special coronary outlet boundary condition, consisting of an RCRCR circuit connected to an intramyocardial pressure, is employed in~\cite{Kim2009}.
Since a direct measurement of intramyocardial pressure is unavailable due to the difficulty of measuring this indicator \emph{in-vivo}, it is often approximated by the left ventricular pressure and its time-derivative \cite{Sankaran2012, Ramachandra2017, Tran2017, Tran2019}. 
However, despite the importance of this quantity in driving coronary flow,  the effect of this approximation is unclear. 
Therefore, in this work, we model the uncertain intramyocardial pressure time-derivative P$_\text{im,t}$ with a stochastic process and assess the corresponding variability in the simulation outputs. 
We first obtained a baseline P$_\text{im,t}$(t) waveform from our previous study~\cite{Tran2017}, and perturbed the intramyocardial pressure time derivative in the systolic phase of the heart cycle using a K-L expansion with process variation $\sigma$ equal to 10\% of the maximum value assumed by P$_\text{im,t}$(t) during the heart cycle. 
Note that variability in the intramyocardial pressure time-derivative is only considered, in this study, during systole (t$<$0.5s) as ventricular contraction determines the diastolic nature of coronary flow. During filling the uncertainty is instead considered negligible.
The perturbed P$_\text{im,t}$(t), and the reconstructed left ventricular pressure P$_{lv}$(t) (up to a constant) are both shown in Figure~\ref{fig:6}(c).

% ====================================================================
\subsection{Uncertainty in the morphometry exponent}\label{sec:morphoUncertainty}
% ====================================================================

\noindent 
The morphometry law defines the mathematical relationship between vessel size and flow rate distribution after vascular branching, $Q\propto r^m$, where $Q$ is the flow rate in a vessel, $r$ is vessel radius, and $m$ is the morphometry exponent. 
Morphometry exponents selected on physiological grounds have been used to assess distal coronary resistance as $R\propto r^{-m}$~\cite{Sankaran2012, Taylor2013, Ramachandra2017, Tran2019}. However, since the exponent $m$ is typically obtained empirically, uncertainty arises when choosing a single value for the exponent in simulations, and there is no quantitative understanding of how variability in $m$ affects simulation results.

A morphometry exponent equal to 3 was first proposed by Murray~\cite{Murray1926}, based on the minimum work principle, where the vessel wall diameter adapts to conserve wall shear stress in bifurcations. 
By analyzing Mall's histological data~\cite{Mall1905}, Sherman~\cite{Sherman1981} validated the cube law ($m=3$), showing that the order of magnitude of the sum of cubed diameters, $\sum r_i^3$ is conserved over the entire vasculature tree, which is consistent with the conservation of mass, $\sum Q_i=\text{const}$. 
While Mayrovitz \etal~\cite{Mayrovitz1983} also validated the cubic dependence for the microvasculature, exponents other than three provided a better fit for larger vessel sizes.
Hutchins \etal~\cite{Hutchins1976} reported $m=3.2\pm1.6$ for the left coronary in the healthy human, while the mean of $m$ decreases to $m=2.2\pm2.1$ with increasing severity of arteriosclerosis, and to $m=2.7\pm1.3$ for other epicardial coronary arteries. 
Additionally, Zamir \etal~\cite{Zamir1992} demonstrated that $m=2$ fits the histological data better than $m=3$ in the major branches of the aortic arch, such as the carotid artery and sinus bifurcation. They argued that the shear stress is not constant for all vessel sizes, and rather increases for smaller vessel sizes, leading to an exponent $m$ that should be less than three.
Based on the above studies, and on the fact that the size of the coronary arteries is smaller than the aortic branches but larger than the microvasculature, a reasonable value of $m$ should lie between $m=2$ and $m=3$. 
Recent studies reported morphometry exponents for the LCA ranging from 2.45 to 2.51~\cite{Zhou1999} and $m=2.6\pm 0.64$ for LCA in~\cite{Changizi2000}.

The total resistance of the downstream vasculature has been determined using various exponents $m$ in coronary hemodynamics literature.
Coronary models in Sankaran \etal~\cite{Sankaran2012} assumed that the resistance is inversely proportional to the area of the distal branch with $m=2$. Taylor \etal~\cite{Taylor2013} used $m=3$ instead for boundary conditions in aorto-coronary models. Jaquet \etal~\cite{Jaquet2017} and other recent studies~\cite{Ramachandra2017, Tran2019} adopted $m=2.6$ for coronary arteries.
In our study, we assume the morphometry exponent as a uniformly distributed random variable, $m(\omega)$ between 2.4 and 2.8, i.e. $m \sim \mathcal{U}(2.4, 2.8)$.

% ===========================================================================
\subsection{Uncertainty in the material properties of vascular tissue}\label{sec:matUncertainty}
% ===========================================================================

\noindent A stochastic model is particularly appropriate to represent inter- and intra- patient variability in the vessel wall elasticity due to the challenges of making direct {\it in-vivo} assessment of vascular tissue histology. 
A study by Gow \etal~\cite{Gow1979} reported the Young's modulus for the femoral artery, coronary artery and aortic arch in nine dogs, leading to several studies applying a Young's modulus equal to 1.15 MPa~\cite{Coogan2013, Ramachandra2017, Tran2019}.
In this study, we use a more recent quantification of elastic modulus, obtained from tensile tests on 13 human coronary vessels, reporting a mean modulus of $\mu[\text{E}_s]=1.48$ MPa and standard deviation of  $\sigma[\text{E}_s]=0.24$ MPa~\cite{Karimi2013}. 
From this data, the vessel Young's modulus is simulated as a spatially uniform Gaussian random variable $\text{E}_s\sim \mathcal{N}(1.48, 0.24^{2})$. Since this distribution is sufficiently bounded away from zero, it coincides closely with a lognormal distribution having the same mean and variance. In addition, we note that all Young's modulus realizations used in this study belong to a physiologically admissible range for coronary arteries, from 0.95MPa to 2.29MPa.

% ==========================================================
\subsection{Uncertainty propagation methodologies}\label{sec:UQdetails}
% ==========================================================
% Overview of uncertainty propagation approaches in the literature
\noindent Several approaches for uncertainty propagation have been proposed in the literature. 
% MC
Monte Carlo sampling (MC) is the simplest method with many advantages over other approaches. Specifically, MC is unbiased, it can handle arbitrary distributed and correlated inputs, its Mean Square Error (MSE) does not depend on the random input dimensionality, and the model evaluations can be solved independently in an embarrassingly parallel fashion.
However its slow convergence rate, MSE$\propto\mathcal{O}(1/\sqrt{N})$, makes this technique cost prohibitive in many applications. 
% QMC
Stratified or low-discrepancy sampling strategies, such as Latin Hypercube or Quasi Monte Carlo sequences (QMC), provide a more uniform coverage of the input domain and have been shown to improve the convergence rate of MC estimators up to $\mathcal{O}(ln N^d/N)$.
Popular examples include the Sobol'~\cite{Sobol1976} or Halton~\cite{Halton1960} sequences, together with techniques which improve the discrepancy in high dimensional settings such as Faure permutations~\cite{faure1992good,Maitre2010}.
Other families of propagation approaches are based on numerical integration.
Stochastic Collocation (SC) methods use multivariate numerical quadrature to compute the expansion coefficients of the system stochastic response according to families of polynomials  orthogonal with respect to the probability measure of the random inputs. These coefficients are then used to compute the statistics of interest.
In addition, the Smolyak algorithm~\cite{Smolyak1963} has been introduced to effectively reduce the computational burden due to the exponential increase in computational cost of evaluating tensor product polynomials or quadrature rules in high dimensions.
However, SC suffers from multiple shortcomings, for example, the difficulty of handling discontinuous response surfaces~\cite{Agarwal2008, Schiavazzi2016}, and the inability to compute the expansion coefficients at the predetermined integration order, when a deterministic solver crashes at a single quadrature point. This precludes SC from being practically usable for a possibly large number of model evaluations. 
More recently, significant research has been devoted to the development of \emph{adaptive} approaches where model evaluations are iteratively placed at appropriate new input realizations, informed by metrics that are progressively refined.
In this context, popular approaches are the multi-element generalized polynomial chaos~\cite{Wan2005}, LARS-based approaches~\cite{blatman2011adaptive} and generalized multi-resolution basis (MW)~\cite{Schiavazzi2014,Schiavazzi2016,Tran2019}. 
Finally, we mention the use of stochastic surrogates based on Gaussian process regression or Kriging~\cite{williams2006gaussian} are recently applied to the solution of forward and inverse problems, including multi-fidelity approaches, in cardiovascular modeling reported, e.g, in~\cite{kissas2020machine}.

In this study, we focus on the MW framework, which is flexible in terms of the locations of the random input samples and well suited to handle response surfaces characterized by sharp gradients or discontinuities. 
In section \ref{sec:results} we evaluate how fast MC, QMC, SC, and MW schemes converge on three simple analytic, and one non-linear benchmark problems. We use the Halton sequence for the QMC, and a sparse Smolyak Clenshaw-Curtis grid for SC. Propagation with MW is computed using both random input samples from integration grids (MW-Grid) and random samples (MW-Random). In MW for propagation in the LCA model, we use the same free parameters and threshold quantities listed in \cite{Schiavazzi2016}. 

% ===================================================================
\subsection{Uncertainty propagation through the LCA model}\label{sec:UQmethods}
% ===================================================================

\noindent We performed 1203 cardiovascular simulations in total, each consisting of four cardiac cycles. 
All simulations were performed in parallel on 48 and 96 cores either using the Comet and the Stampede cluster available through the NSF-funded XSEDE portal, or using the Notre-Dame Center for Research Computing cluster.
The parallel scalability has been tested and the number of mesh elements per core used in this study shown to achieve excellent efficiency~\cite{Seo2019}. 
For all simulations we confirmed that the transient effect from the initial condition was eliminated after two cardiac cycles, and we post-processed data only in the last two cycles. 
Both QMC Sobol' sequences and sparse grid stochastic collocation were used to produce input parameter realizations.
In each P$_\text{in}$ and P$_\text{im,t}$ study, we use four dimensional independent random variables to represent the perturbed pressure waveforms using the K-L expansion. For each E$_s$ and $m$, we use one dimensional random variable. Lastly, we perturbed all quantities of interest simultaneously, resulting in a total of 10 random inputs. 
Using the QMC sampling, we performed 200 simulations for the inlet pressure stochastic process P$_\text{in}$, 200 simulations for the intramyocardial pressure P$_\text{im}$, 100 simulations for the Young's modulus E$_s$ and 100 simulations for the morphometry exponent, $m$. We then performed 200 simulations with all input parameteres perturbed, including P$_\text{in}$, P$_\text{im}$, $m$, and $E$.
Additionally, we ran 137 simulations for P$_\text{in}$, 137 simulations for P$_\text{im}$, and 129 simulations for E$_s$, all from a Smolyak sparse grid~\cite{Smolyak1963}. 

We focused on four quantities of interests (QoI) related to both hemodynamics and wall mechanics. 
We calculated the branch flow rate Q$_i$(t), pressure P$_i$(t), time-averaged wall shear stress TAWSS$_i$(x), wall deformation Deformation$_i$(x) in all six branches ($i=1,...,n_o$).
The pressure and flow is area-averaged at each branch outlet, while the shear stress, $\vec{\tau}({\bf x},t)$ and wall displacement $\vec{\delta}({\bf x},t)$ are time-averaged and spatially averaged over the vessels circumference, i.e. 
\beq
\text{TAWSS}_i\left(\frac{x}{L_i}\right)=\int_{C}\frac{1}{\text{T}}\left|\int_{0}^\text{T} \vec{\tau}({\bf x},t)\,dt\right| ds,\;\;\text{Deformation}_i
\left(\frac{x}{L_i}\right)=\int_{C}\frac{1}{\text{T}}\left|\int_{0}^\text{T} \vec{\delta}({\bf x},t)\,dt\right| ds, 
\eeq
where $\text{T}$ is the cycle duration, $t$ is time, $x$ is the spatial distance along the centerline of the vessel branch, $L_i$ is the axial length of a vessel branch, and $C$ the circumference of the slice plane perpendicular to the centerline.

% ==============================================
\section{Uncertainty propagation results}\label{sec:results}
% ==============================================

% ==========================================================
\subsection{Benchmark: continuous and discontinuous analytic functions}
% ==========================================================

\begin{figure}[ht!]
\vspace{-6pt}
\centering
\includegraphics[width=0.78\textwidth, keepaspectratio]{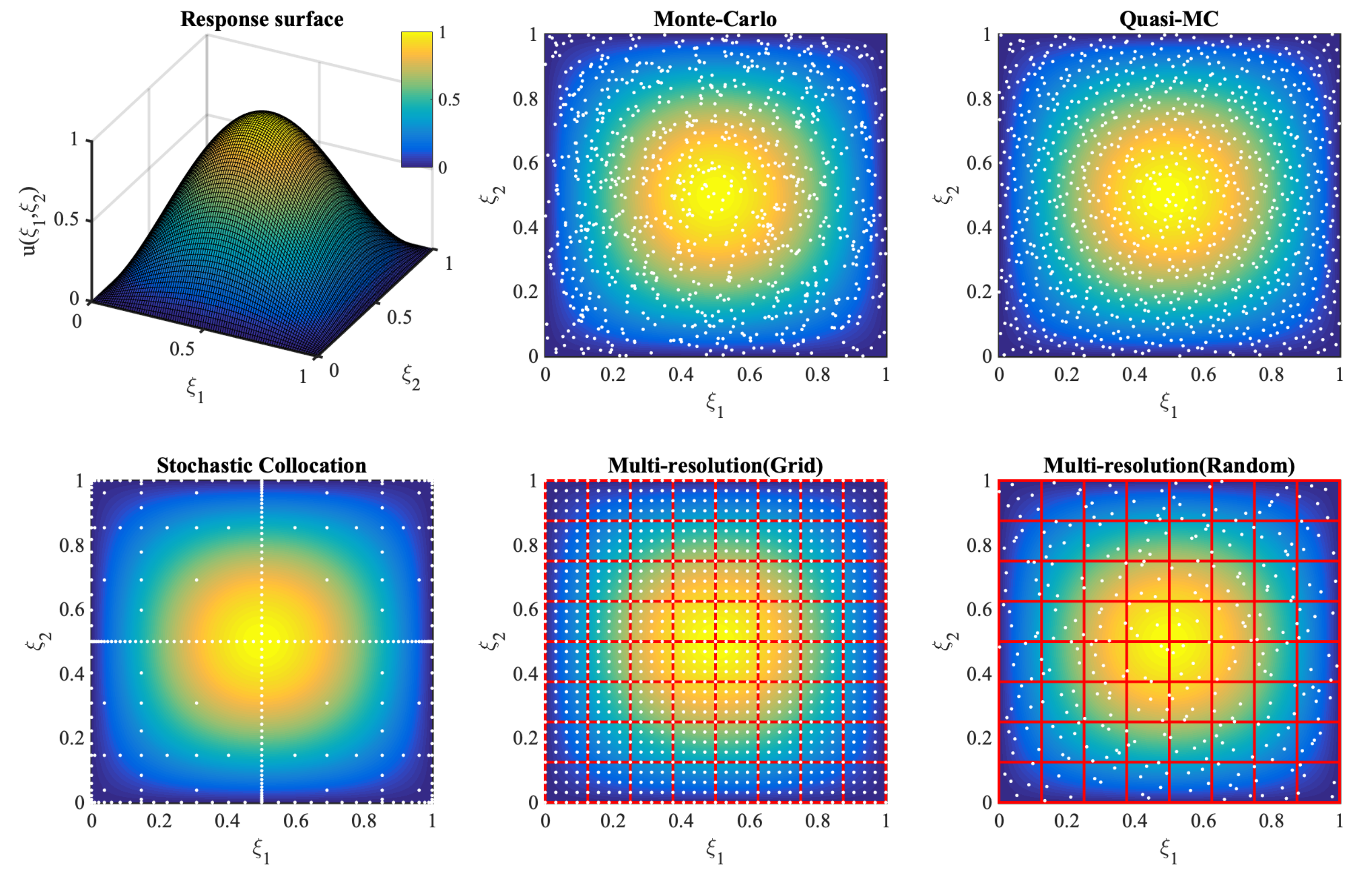}
\includegraphics[width=0.75\textwidth, keepaspectratio]{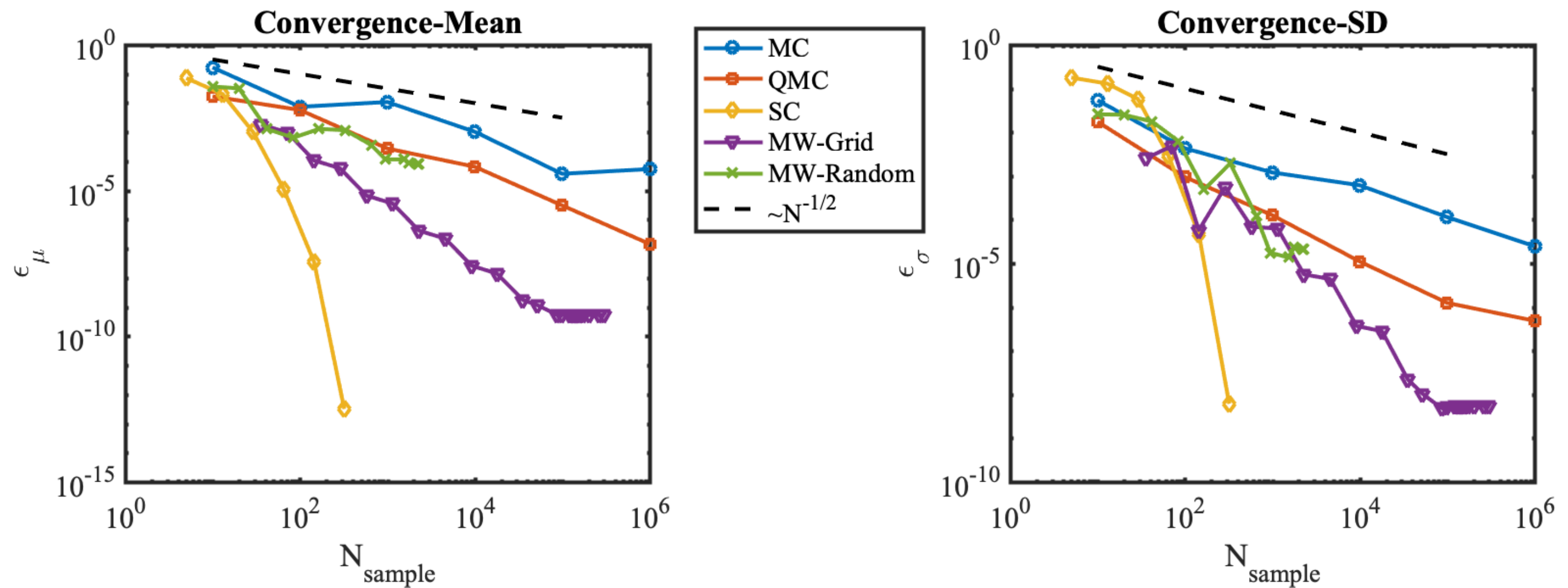}
\mylab{-125mm}{127.5mm}{(a)}%
\mylab{-83mm}{127.5mm}{(b)}%
\mylab{-40mm}{127.5mm}{(c)}%
\mylab{-125mm}{86mm}{(d)}%
\mylab{-83mm}{86mm}{(e)}%
\mylab{-40mm}{86mm}{(f)}%
\mylab{-123mm}{45mm}{(g)}%
\mylab{-52mm}{45mm}{(h)}%
\caption{Model response and sampling distributions in forward uncertainty propagation. (a) The response surface of $u(\xi_1, \xi_2)=sin(\pi\xi_1)sin(\pi\xi_2)$. (b-f) Sampling points (white dots) are overlaid on the response surface (colormap). (b) Monte-Carlo, (c) Quasi Monte-Carlo, (d) Clenshaw-Curtis sparse grid with level 6, (e) Multi-resolution framework using grid points. The hypercube binary partitions are plotted as red boxes, at refinement iteration 7. (f) Multi-resolution framework using random points, at refinement iteration 12.  (g) Absolute errors in the mean, $\epsilon_\mu$, (h) absolute errors in the standard deviation.
}\label{fig:7}
\end{figure}

\noindent We first tested several UQ methodologies on a sinusoidal response surface with two-dimensional random variables, $\xi_1$ and $\xi_2$, i.e.,
\begin{equation}
  u(\xi_1,\xi_2)=      sin(\pi\xi_1)sin(\pi\xi_2), 
      \label{eq:SW}
\end{equation}
in which $\xi_1$ and $\xi_2$ are uniform random variables ranging between 0 and 1. 
Samples from MC, QMC, SC, and MW methods are plotted on the sinusoidal response surface in Figure~\ref{fig:7}, respectively. 
In terms of convergence, SC is outperforming all methods (Figure~\ref{fig:7}(g)), resulting, as expected, in excellent accuracy on smooth response surfaces.
For this benchmark, we confirm the superiority of low-discrepancy QMC sequences (Figure~\ref{fig:7}(c)) over standard Monte Carlo. MW with random samples led to comparable performance against QMC, while MW with the quadrature grid showed superior performance against QMC.
\begin{figure}[t]
\vspace{-6pt}
\centering
\includegraphics[width=0.8\textwidth, keepaspectratio]{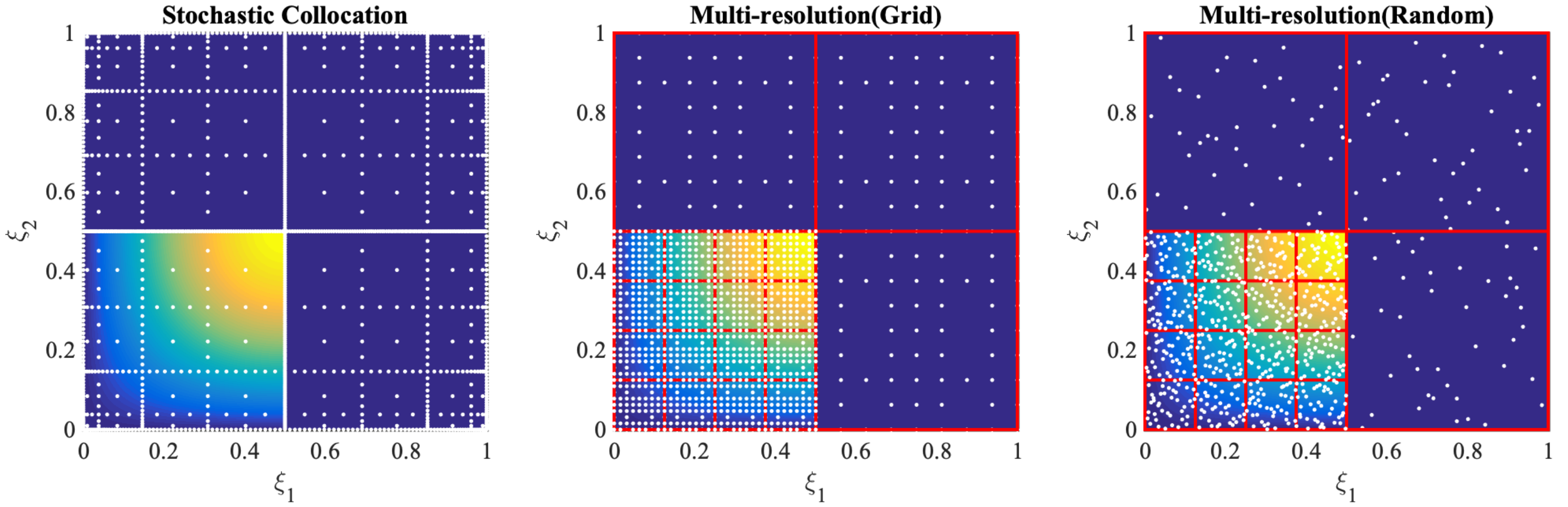}
\includegraphics[width=0.7\textwidth, keepaspectratio]{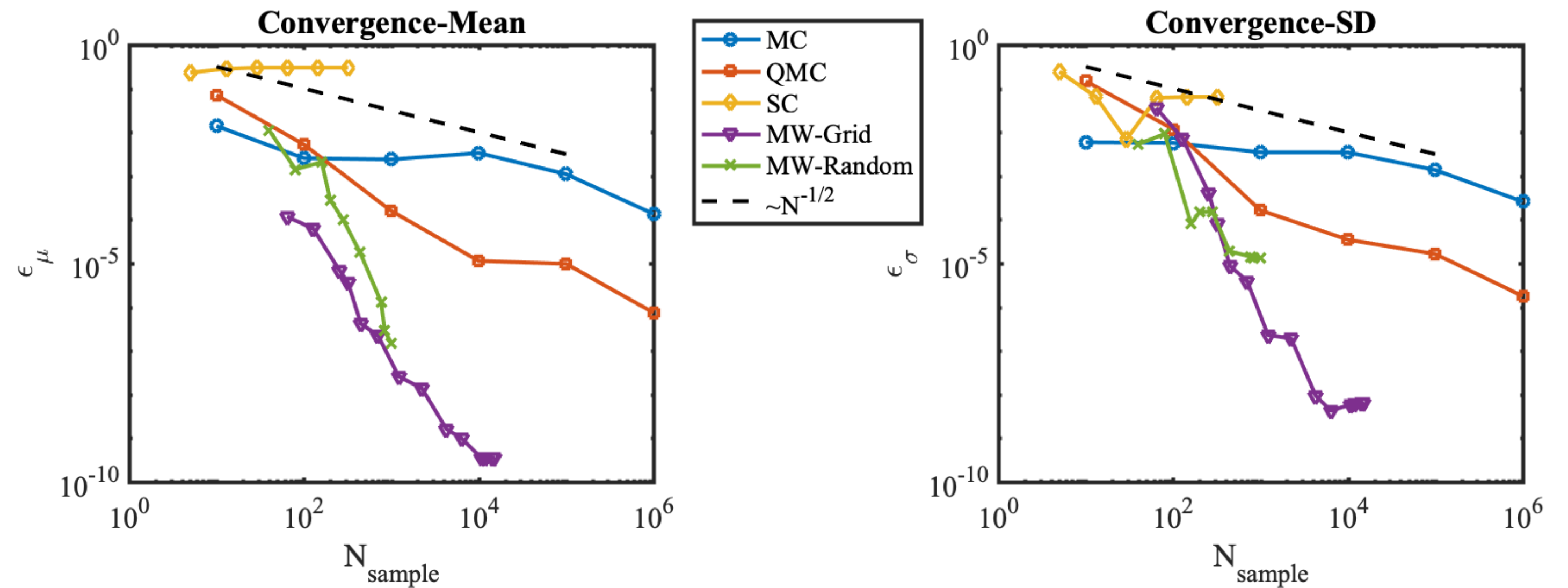}
\mylab{-124mm}{85mm}{(a)}%
\mylab{-80mm}{85mm}{(b)}%
\mylab{-35mm}{85mm}{(c)}%
\mylab{-115mm}{41mm}{(d)}%
\mylab{-48mm}{42mm}{(e)}%
\caption{Sampling methods and convergence rates for the discontinuous sine response surface from equation \eqref{eq:TS}. Sampling points (white dots) are overlaid on the response surface (colormap). (a) Sparse Clenshaw-Curtis grid with the Smolyak, level 8, (b) Multi-resolution framework using grid points. The hypercube binary partitions are plotted as red boxes, at refinement iteration 7. (c) Multi-resolution framework using random points, at refinement iteration 13. 40 initial samples were placed. (d) Absolute errors of the mean, $\epsilon_\mu$, (e) absolute errors of the standard deviation, $\epsilon_\sigma$.}\label{fig:8}
\end{figure} 
Next, we consider a truncated sine surface,  
\begin{equation}
  u(\xi_1,\xi_2)=
    \begin{cases}
      sin(\pi\xi_1)sin(\pi\xi_2) &\text{if}\;(\xi_1\leq 0.5,\;\; \xi_2\leq 0.5)\\
      0 & \text{otherwise}
    \end{cases},
    \label{eq:TS}
\end{equation}
which is positive in the third quadrant of the stochastic domain, and zero elsewhere. 
In this case, the performance of SC deteriorates as shown in Figure~\ref{fig:8}, and shown previously in literature for problems with stochastic responses characterized by sharp gradients and discontinuities~\cite{Agarwal2008}. 
QMC shows faster convergence than MC. MW, especially MW-Grid, substantially outperforms all other methods on this second benchmark.

\begin{figure}[t]
\vspace{-6pt}
\centering
\includegraphics[width=1.0\textwidth, keepaspectratio]{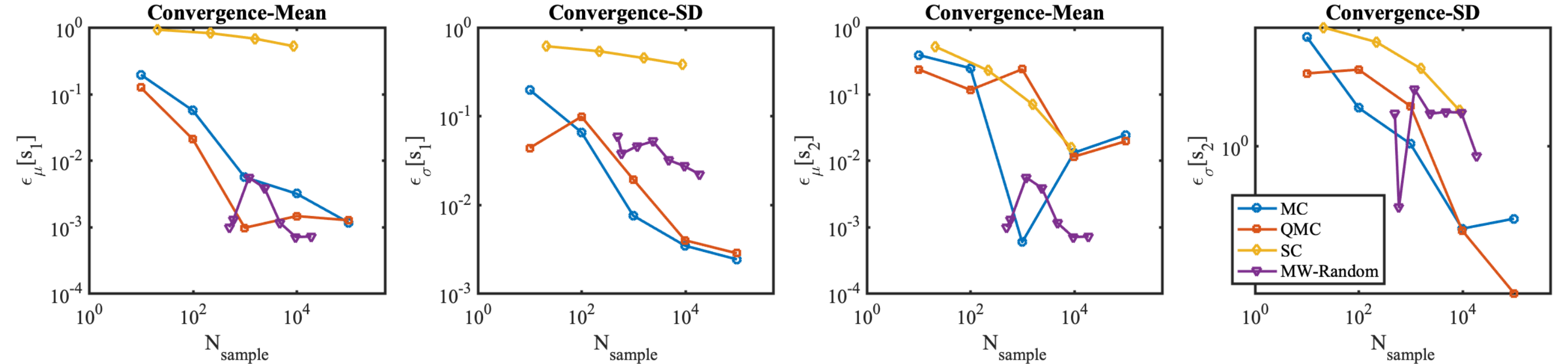}
\mylab{-81mm}{41mm}{(a)}%
\mylab{-40mm}{41mm}{(b)}%
\mylab{1mm}{41mm}{(c)}%
\mylab{44mm}{41mm}{(d)}%
\caption{Convergence rates for the ten dimensional sine response surface from equation \eqref{eq:Sobol}. (a) Absolute errors of the mean, $\epsilon_\mu[s_1]$, (b) absolute errors of the standard deviation, $\epsilon_\sigma[s_1]$, (c) absolute errors of the mean, $\epsilon_\mu[s_2]$, (d) absolute errors of the standard deviation, $\epsilon_\sigma[s_2]$.}\label{fig:9}
\end{figure}

In addition, we performed uncertainty propagation using two high-dimensional response functions. We selected the two Sobol' functions \cite{chen2015new}:
\begin{equation}
s_{1}(\boldsymbol{\xi}) = \prod_{i=1}^{d}\,\frac{\vert 4\,\xi_{i} - 2\vert + p_{i}}{1 + p_{i}},\,\,\text{and}\,\,s_{2}(\boldsymbol{\xi}) = \prod_{i=1}^{d}\,\frac{1 + 3\,p_{i}\,y^{2}_{i}}{1 + p_{i}},
    \label{eq:Sobol}
\end{equation}
where $\boldsymbol{\xi}\in[0,1]^{d}$ and $\mathbf{p} = (p_{1},p_{2},\dots,p_{d})$ is a vector of non-negative and positive parameters for $s_{1}(\boldsymbol{\xi})$ and $s_{2}(\boldsymbol{\xi})$, respectively. Here we select $d=10$ and $p_{i}=2,\,i=1,\dots,d$ for both functions. Integration of the above functions with respect to the uniform measure in $[0,1]^{d}$ leads to:
\begin{equation}
\mu[s_{1}] = 1,\,\,\text{and}\,\,\sigma[s_{1}] = \sqrt{\left[\prod_{i=1}^{d}\,\left(\frac{1}{1 + p_{i}}\right)^{2}\,\left(\frac{4}{3} + p^{2}_{i} + 2\,p_{i}\right)\right] - 1},
    \label{eq:S1}
\end{equation}
and
\begin{equation}
\mu[s_{2}] = 1,\,\,\text{and}\,\,\sigma[s_{2}] = \sqrt{\left[\prod_{i=1}^{d}\,\left(\frac{1}{1 + p_{i}}\right)^{2}\,\left(\frac{9}{5}\,p^{2}_{i} + 1 + 2\,p_{i}\right)\right] - 1}.
    \label{eq:S2}
\end{equation}
Figure ~\ref{fig:9} illustrates the efficiency of several methods in computing the mean and standard deviation of these functions. SC performs poorly in this case, even though the response surface is not discontinuous. QMC and MC performed almost equally fast. MW-Random shows the superior performance for estimation of means, especially in low number of samples around 100.

% =================================================
\subsection{Benchmark: 2-D Kraichnan-Orzag}\label{sec:KO2D}
% =================================================

\begin{figure}[ht]
\vspace{-6pt}
\centering
\includegraphics[width=0.9\textwidth, keepaspectratio]{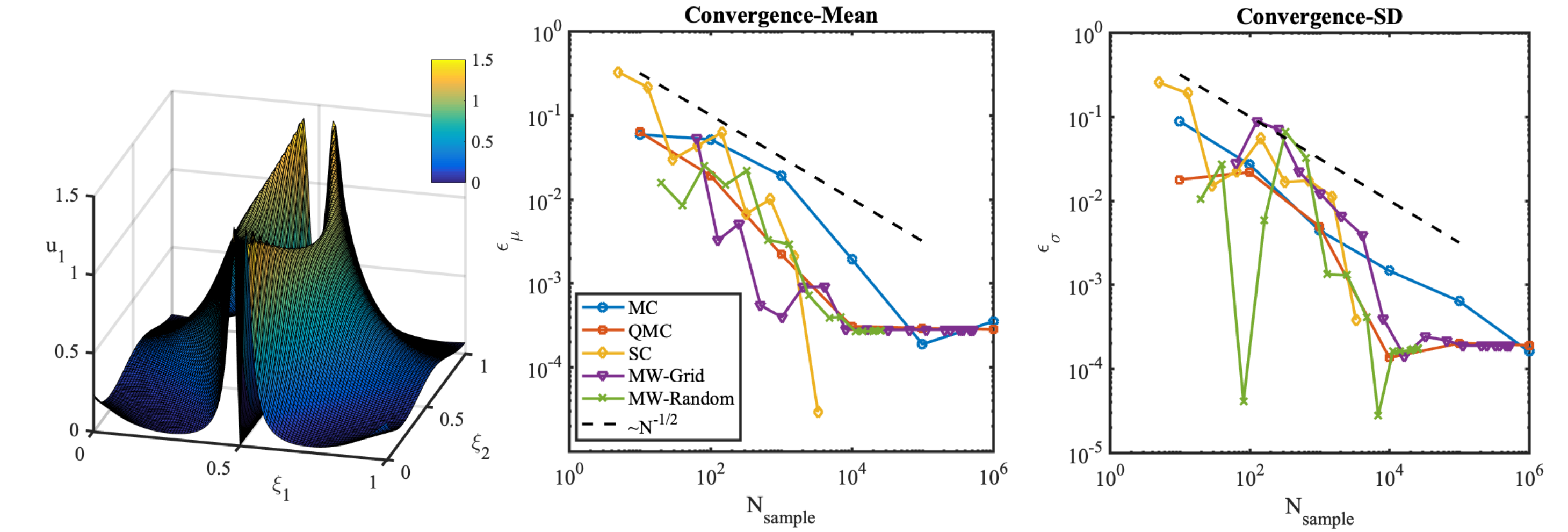}
\mylab{-146mm}{47mm}{(a)}%
\mylab{-105mm}{47mm}{(b)}%
\mylab{-52mm}{47mm}{(c)}%
\caption{Uncertainty propagation through the Kraichnan-Orzag problem. (a) the response surface of $u_1$ at t=10 subject to uncertain parameters in the initial condition $\xi_1$ and $\xi_2$. (b) the error of the mean of $u_1$ at t=10, $\epsilon_\mu$, against the number of samples. (c) the error of the standard deviation of $u_1$ at t=10, $\epsilon_\sigma$, against the number of samples. The reference mean and standard deviation are obtained from MC with $2\times 10^6$ samples.}\label{fig:10}
\end{figure}

\noindent The performance comparison continues with the Kraichnan-Orszag (K-O) system, a well known benchmark for adaptive propagation. This is a differential system consisting of three equations
\begin{equation}\label{eq:KO}
\frac{du_1}{dt}=u_1u_3,\quad\frac{du_2}{dt}=-u_2u_3,\quad\frac{du_3}{dt}=u_1^2+u_2^2,
\end{equation}
which is a non-linear, three dimensional, time-dependent differential system derived, by truncation, from the Navier-Stokes equations~\cite{Kraichnan1963, Orszag1967}. 
The initial condition centered at $(u_1,u_2,u_3)=(1,0,0)$ is perturbed in two directions as
\begin{equation}\label{eq:IC}
u_1(t=0)=1,\quad u_2(t=0)=0.2\xi_1-0.1, \quad u_3(t=0)=2\xi_2-1, 
\end{equation}
in which $\xi_1$ and $\xi_2$ are uniform random variables with range in $[0,1]$. 
This perturbation significantly alters the solution, particularly at large integration times with a resulting stochastic response at time 10 characterized by sharp edges as shown in Figure~\ref{fig:10}(a). 
As shown in Figure~\ref{fig:10}(b) and Figure~\ref{fig:10}(c), the performance of QMC is superior to MC, as expected. SC shows slower convergence than QMC below 1000 samples, but faster with additional samples and quickly converge after 1000 samples.
MW-Grid shows good accuracy in computing the response mean, especially with a small number of samples, but with a larger standard deviation than MC. MW-Random exhibits superior performance for both the mean and standard deviation.

From the three benchmarks above, we confirm the superiority of QMC with respect to MC. 
Also we observe outstanding performance of SC on a smooth response surface, but confirm that this approach is not designed to handle discontinuities or sharp gradients. 
MW shows faster convergence than QMC in most cases, and outperforms SC in problems with non-smooth response.
Based on these commulative results, we chose to apply QMC (with Sobol' sequences), SC and MW forward propagation to our left coronary artery model.
Note that adaptivity in MW is restricted only to the input domain partitions, keeping the number of samples in the QMC sequence fixed, or, in other words, no samples are incrementally added with respect to the QMC sequences or quadrature grids selected \emph{a priori}.

% ==========================================================
\subsection{LCA: Effects of uncertainty in the inlet pressure, P$_{in}$}\label{sec:uqPress}
% ==========================================================

\begin{figure}[ht!]
\vspace{-6pt}
\centering
\includegraphics[width=0.9\textwidth, keepaspectratio]{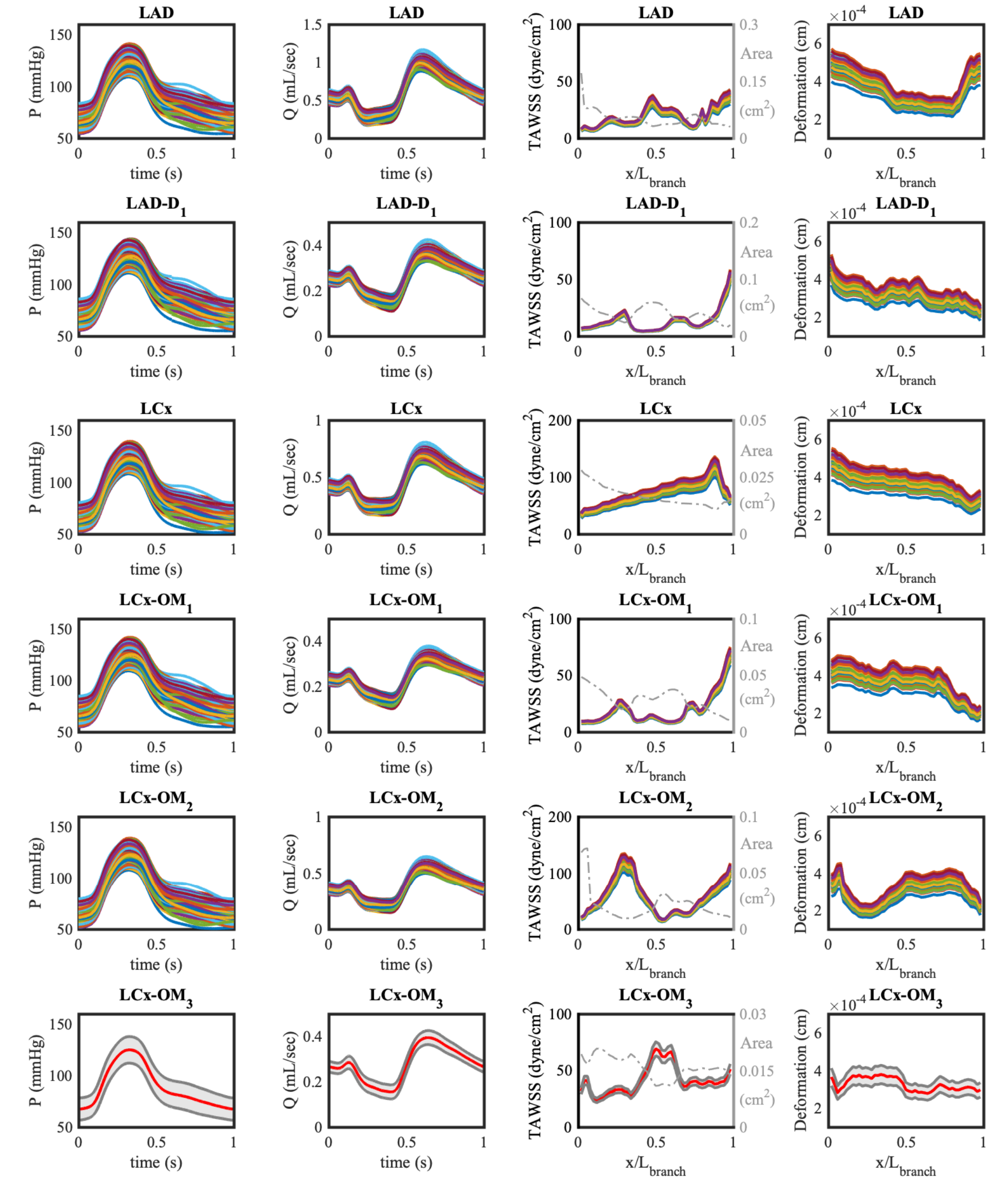}
\caption{200 realizations of QoI for six branches of LCA model with the perturbed pulsatile inlet pressure. Realizations are computed from the QMC sampling.  The cross-sectional area of vessel branch is plotted on top of TAWSS with a dashed dot gray line. The ensemble averaged quantities of the LC$x$-OM$_3$ branch are plotted with red lines, and 95 percent confidence intervals are plotted in gray. $x$ is the axial location along the centerline of the vessel, and L$_\text{branch}$ is the length of each branch. }\label{fig:11}
\end{figure}

\noindent Figure \ref{fig:11} shows 200 realizations of waveforms for the output QoIs in six left coronary branches resulting from perturbing the inlet pressure, and indicating all the QoIs at all branches to be significantly affected by this perturbation. 

\begin{figure}[t]
\vspace{-6pt}
\centering
\includegraphics[width=0.95\textwidth, keepaspectratio]{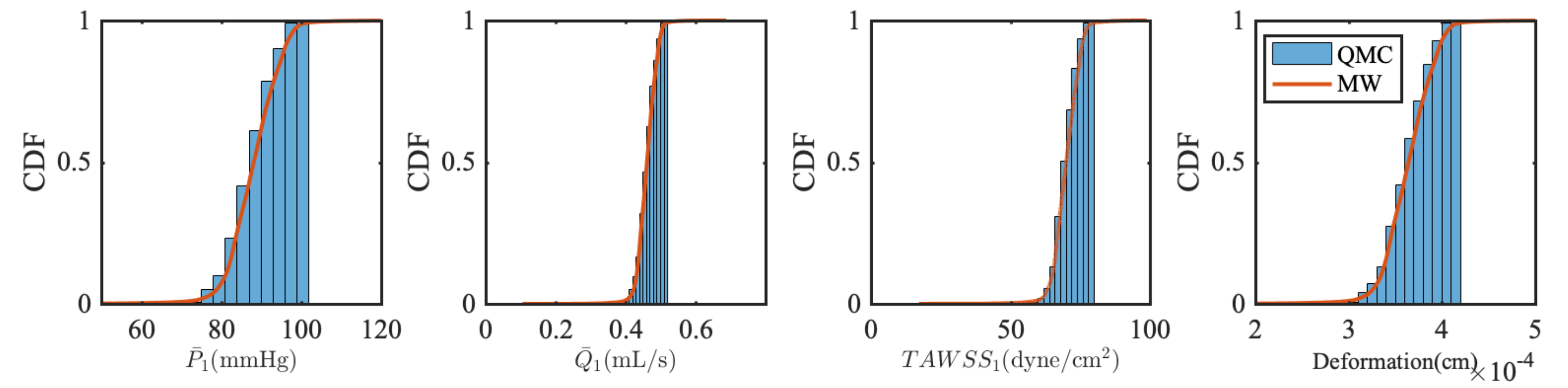}
\includegraphics[width=0.9\textwidth, keepaspectratio]{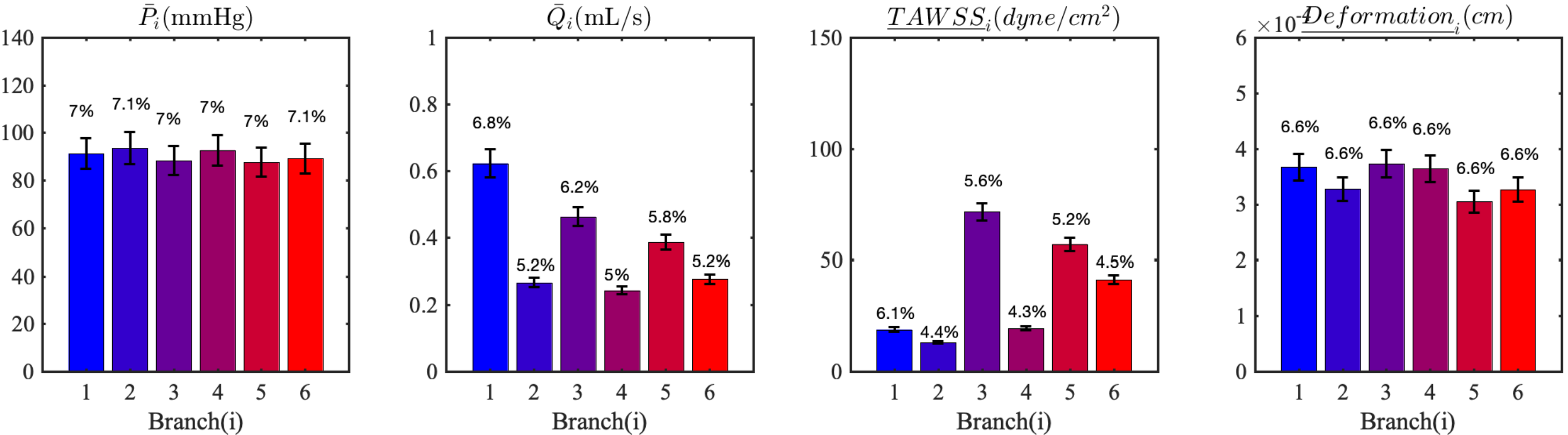}
\caption{(Top) CDF estimates of QoIs in LAD obtained from QMC and MW approach. (bottom) Mean and standard deviations of QoIs resulting from uncertainty in the inlet pressure time history. The error bars represent two standard deviations. The coefficients of variation are reported at the top of each bar. Over bar notation on QoIs means time-average and underline notation means the spatial average.}\label{fig:12}
\end{figure}

Additionally, we quantified Cumulative Distribution Function (CDF) estimates, the mean, standard deviation and coefficient of variation of all QoIs in the six branches (Figure~\ref{fig:12}). 
% Outlet pressure is characterized by the same CoV
A consistent $\overline{c}_v$ equal to 7\% in the outlet pressures at all branches is directly induced by 7\% $\overline{c}_v$ in the variability of the inlet pressure. In our model, only viscous resistance at the wall plays a contributing role for the pressure drop from the inlet to the outlets of the model, since the model is free from sudden geometric variations in diameter, that may induce a significant pressure drop (for example, stenosis). 
In this case, the outlet pressure is related to the inlet pressure through the relation, P$_\text{outlet}$(t)=P$_\text{in}(t)-$RQ(t), where R(dyne$\cdot$s/cm$^5$) is viscous resistance of a vessel branch and Q(t)(mL/s) is corresponding the flow rate. 
Since Q(t) is mainly determined by the pressure difference between P$_\text{im}$(t) and P$_\text{in}$(t), and R is independent of the pressure ($8\mu L / \pi r^4$ for the Poiseuille flow), changes in the inlet pressure directly translate to pressure changes at the outlets. 

% Wall deformation has the same CoV
The displacement of the wall also exhibits the same coefficient of variation prescribed for the inlet pressure, 7\% $\overline{c}_v$. 
In this context, we recall the expression of the radial displacement $\delta_{r}$ of an ideal thick-walled cylinder subject to an internal pressure $p$
\begin{equation}
\delta_{r} = \frac{p\,r_{i}}{E}\,\left(\frac{r^{2}_{o} + r^{2}_{i}}{r^{2}_{o} - r^{2}_{i}} + \nu\right),
\label{eq:walldef}
\end{equation}
where $E$ is the elastic modulus, $r_{i}$ and $r_{o}$ are the inner and outer diameter, respectively and $\nu$ is the Poisson ratio. Although our deformation measures the displacement in any direction, $\delta_{r}\propto p$ provides a good explanation for the linear transformation of the 7\% $c_{v}$ from inlet pressure to the 7\% $\overline{c}_v$ in the wall radial displacement. 

% Flow rates and TAWSS
Flow rates and TAWSS are instead characterized by coefficients of variation slightly less then 7\%. Particularly the branches with higher flow rate have larger coefficient of variation, for example, $i=1, 3$. This means that the elevation of inlet pressure has a greater effect on the longer major coronary artery branches with larger flow rate, while the smaller bifurcated branches are less affected by inlet pressure changes. 
The wall shear stress in Poiseuille flow is characterized by the expression
\begin{equation}
|\tau| = \frac{4\,\mu\,Q}{\pi\,r^{3}_{i}}
\label{eq:twass}
\end{equation}
where $Q$ is the instantaneous flow rate and $\mu$ the dynamic viscosity. 
Therefore, the proportionality between $\tau$ and $Q$ justifies the similar $c_{v}$ obtained for both flow rate and wall shear stress. 
In later sections with additional propagation studies, we will show that the $\overline{c}_v$ in $p$ and $\delta$ are closely related, and $\overline{c}_v$ in $Q$ and TAWSS are directly associated, consistent with the above equations~\eqref{eq:walldef} and~\eqref{eq:twass}.

Lastly, we note the inverse relationship between the variations of cross-sectional area and variations of TAWSS along the axial location in a vessel branch. 
In figure \ref{fig:11}, TAWSS has high peaks where the cross sectional area is small, and low peaks where the cross sectional area is large. 
This is consistent with prior findings reported for coronary bypass grafts~\cite{Tran2019}, and the fact that the morphometry exponent is less than 3 for coronary arteries. 
In contrast, the wall deformation curve along the axial location shows similar trends to the area change. 
The linear relationship between inner wall diameter and the radial displacement in~\eqref{eq:walldef}, assuming the outer diameter is proportional to the inner diameter, is consistent with our finding. 
%

% ===================================================================
\subsection{LCA: Effects of uncertainty in the intramyocardial pressure, P$_{im,t}$}\label{sec:uqIntra}
% ===================================================================

\noindent Next, we perturb the time derivative of the intramyocardial pressure, P$_\text{im,t}$(t), through a K-L expansion, using a standard deviation equal to 10\% of the maximum absolute value in the baseline curve (see Figure~\ref{fig:6}(c)).
In our coronary model, a pressure Neumann boundary condition is imposed at the inlet and the flow is driven by the pressure differences between the inlet and the downstream pressure. 
This way, a perturbation in the intra-myocardial pressure will directly affect the pressure difference between inlet and outlets and hence affect the flow. This contrasts with other approaches in the literature where the inlet flow is prescribed as a part of open loop boundary conditions.

\begin{figure}[t]
\vspace{-6pt}
\centering
\includegraphics[width=0.7\textwidth, keepaspectratio]{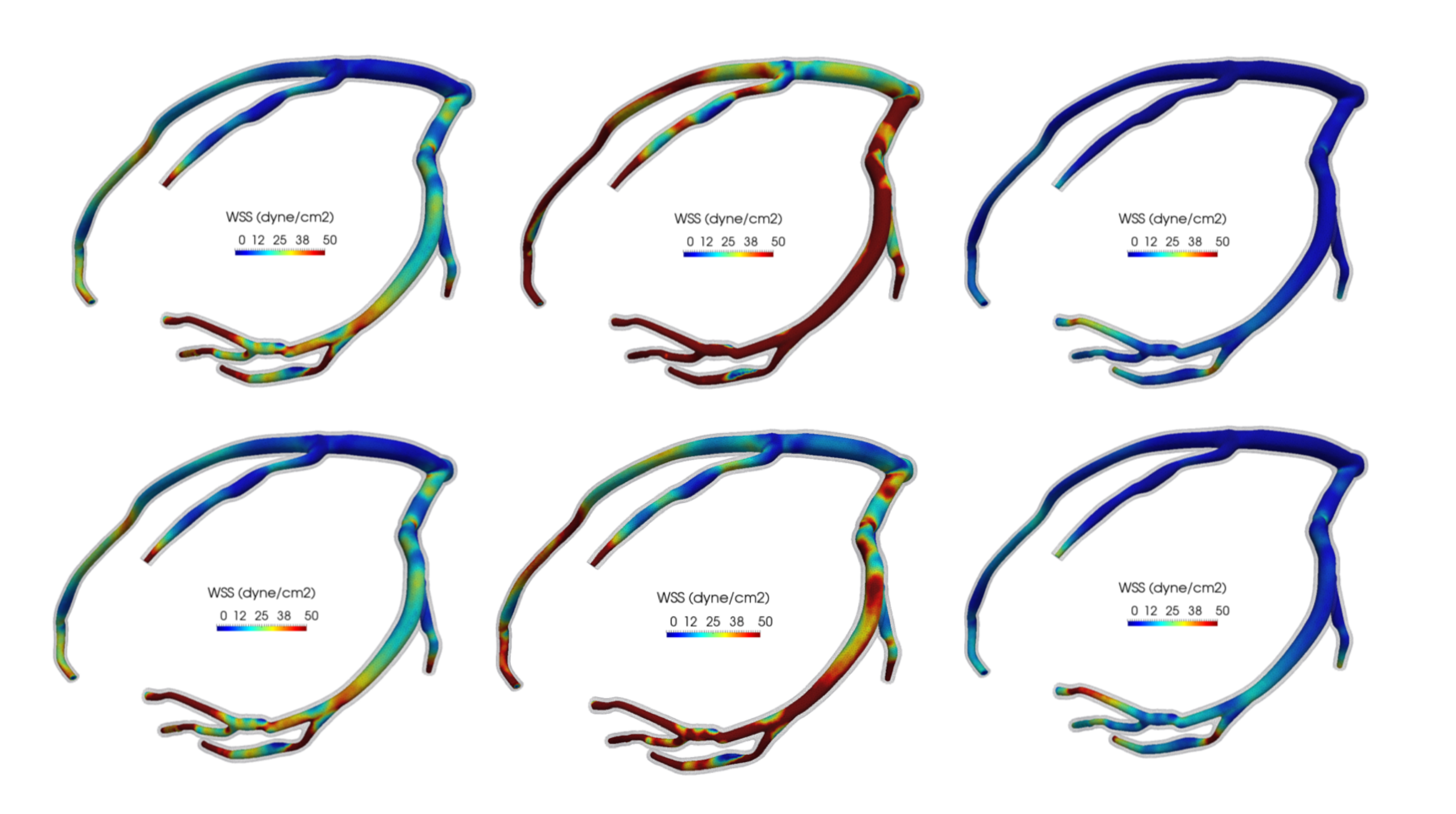}
\caption{Six realizations of the wall shear stress contours on the LCA model resulting from uncertainty in the intramyocardial pressure time history. The wall mesh is colored with transparent gray.}\label{fig:13}
\end{figure}

\begin{figure}[t]
\vspace{-6pt} 
\centering
\includegraphics[width=0.9\textwidth, keepaspectratio]{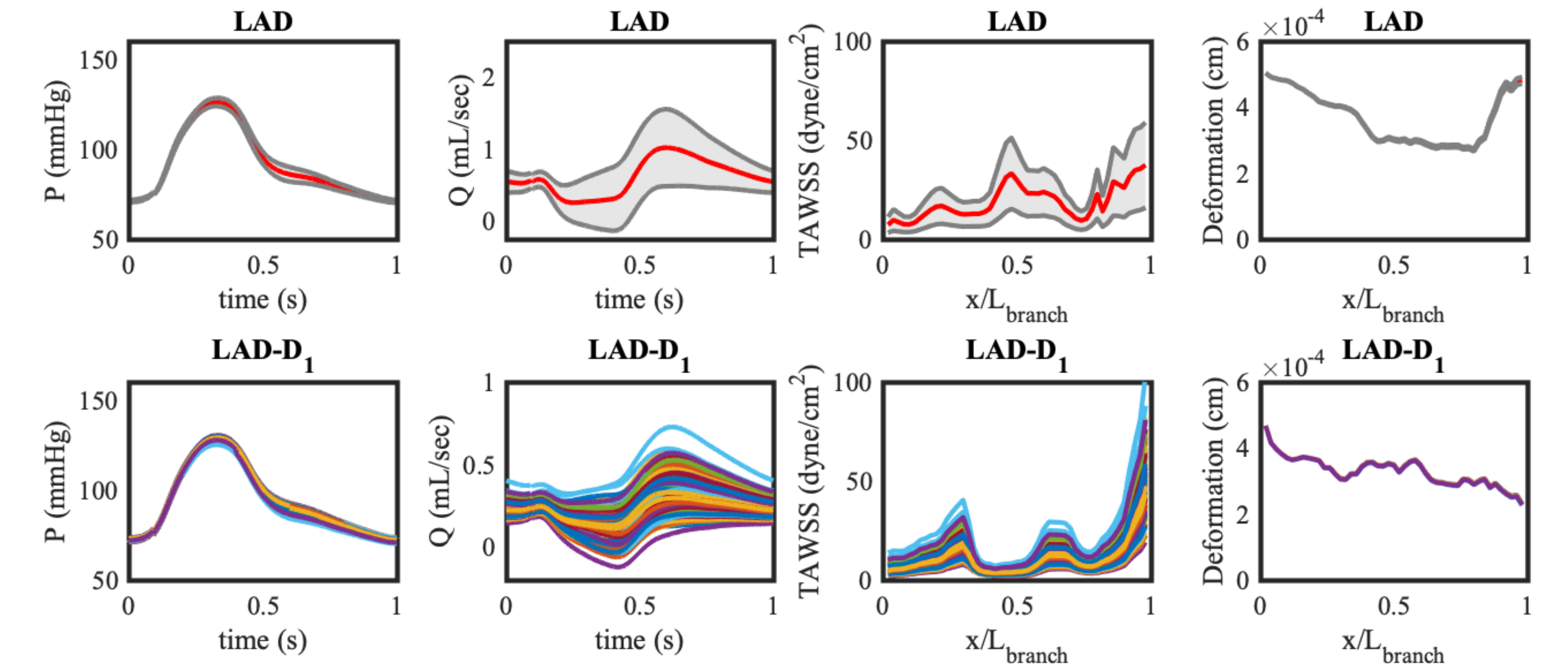}
\includegraphics[width=0.9\textwidth, keepaspectratio]{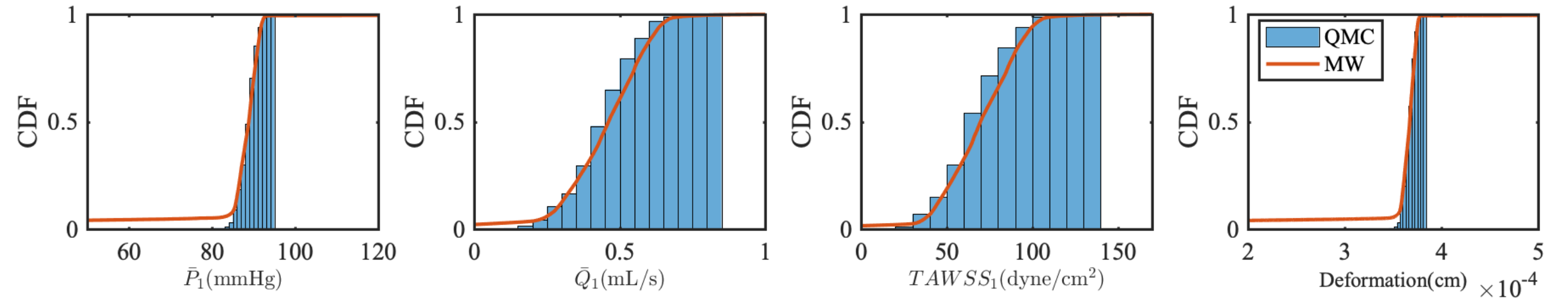}
\includegraphics[width=0.85\textwidth, keepaspectratio]{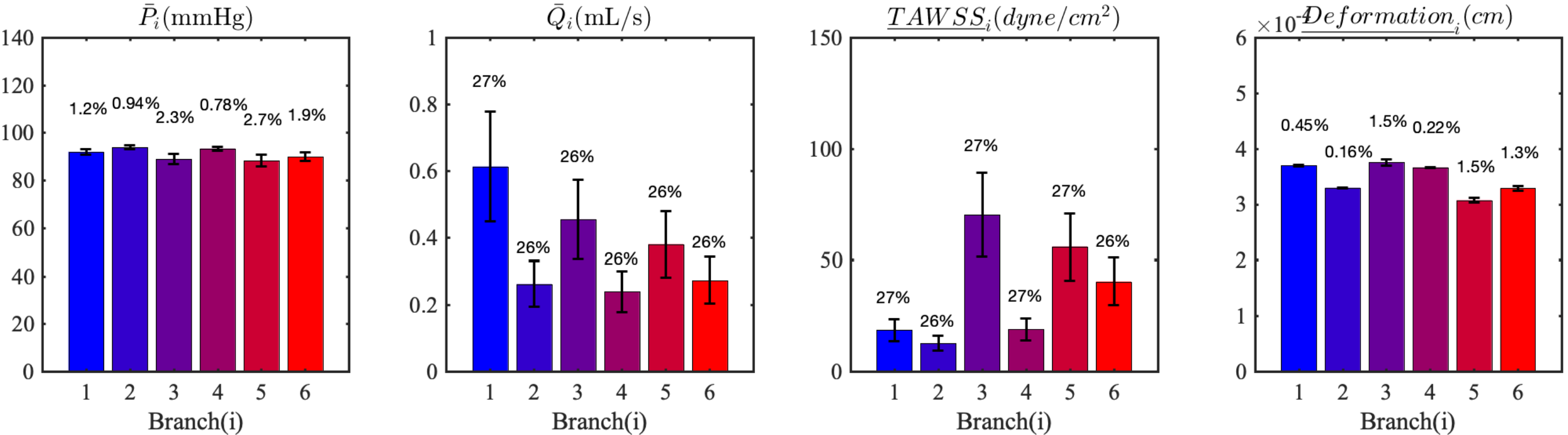}
\caption{Realizations and statistics of the QoIs in two branches resulting from the perturbed time derivative of intramyocardial pressure. In LAD, ensemble averaged quantities are plotted with red lines, and 95 percent confidence intervals are plotted in gray. In the LAD-D$_1$ branch, 200 realizations of QoI are plotted. (Histograms) CDF estimates of QoIs in the LAD branch. (Bar plots) Mean and standard deviations of LCA QoIs resulting from uncertainty in the intramyocardial pressure time history. The coefficients of variation are reported at the top of each bar.}\label{fig:14}
\end{figure}

The results indicate that a 10\% change in P$_\text{im,t}$(t) leads to significant change in TAWSS and flow rate, up to 27 \% $\overline{c}_v$, on the left coronary model (Figures ~\ref{fig:13} and~\ref{fig:14}). 

The coefficient of variation and error bars for all branches are summarized in Figure~\ref{fig:14}. 
The results show a remarkable similarity between the variability in the flow rate $Q$ and TAWSS, and between the variability in pressure and vessel wall deformation, respectively. 

Increasing flow rate is the main source of increase in TAWSS, and the degree of variability ($\overline{c}_v$) in TAWSS is directly transferred to the variability of flow rate. On the other hand, the outlet pressures and deformations are not significantly altered. 
Figure~\ref{fig:14} shows $\overline{c}_v$ for pressure and deformation less than 3\%. As discussed in the previous section, the outlet pressure is directly translated from the fixed inlet pressure, while changes in the downstream intra-myocardial pressure do not significantly affect the pressures at the outlet. 
Since the intra-luminal pressure is the major source of the wall deformation, the wall displacement variability is low, of the same order than the variability in the pressure. 

% ====================================================================
\subsection{LCA: Effects of uncertainty in the morphometry exponent, $m$}\label{sec:uqMorpho}
% ====================================================================

\begin{figure}[ht]
\vspace{-6pt}
\centering
\includegraphics[width=0.9\textwidth, keepaspectratio]{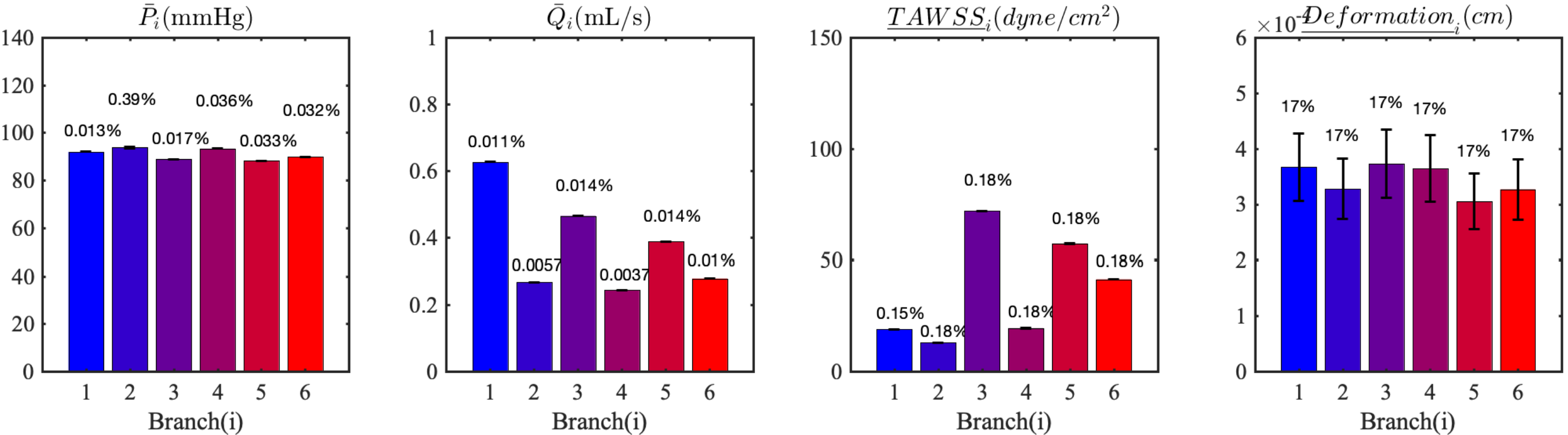}
\caption{Mean and standard deviations of LCA QoI resulting from uncertainty in the Young's modulus. The error bars represent two standard deviation. The coefficients of variation are reported at the top of each bar.}\label{fig:16}
\end{figure}

\noindent The morphometry exponent $m$ is sampled from the uniform distribution $\mathcal{U}\sim (2.4,2.8)$, and used to distribute the total downstream resistances among outlets. 
The results of the propagation, together with the outlet area and the resistances, are plotted in Figure~\ref{fig:16}. 
Perturbation of the morphometry exponent produces changes in the distribution of the vascular resistances, R$_i$ in the six branches, leading up to 4 percent $\overline{c}_v$ in resulting R$_i$. 
The variability of flow rate is reduced from the 4 percent $\overline{c}_v$ in $R_i$ to a maximum $\overline{c}_v$ less than 2$\%$ for Q$_i$, whereas other QoIs exhibit negligible changes. 
With fixed model outlet areas, the variations in the morphometry constant produces a little change in the downstream resistances and the variability on the flow rate is further reduced. 
From the results in this section it can be concluded that, the range of morphometry exponents considered in our study does not lead to sensible effects in the simulation outputs.
\begin{figure}[ht]
\vspace{-6pt}
\centering
\includegraphics[width=0.85\textwidth, keepaspectratio]{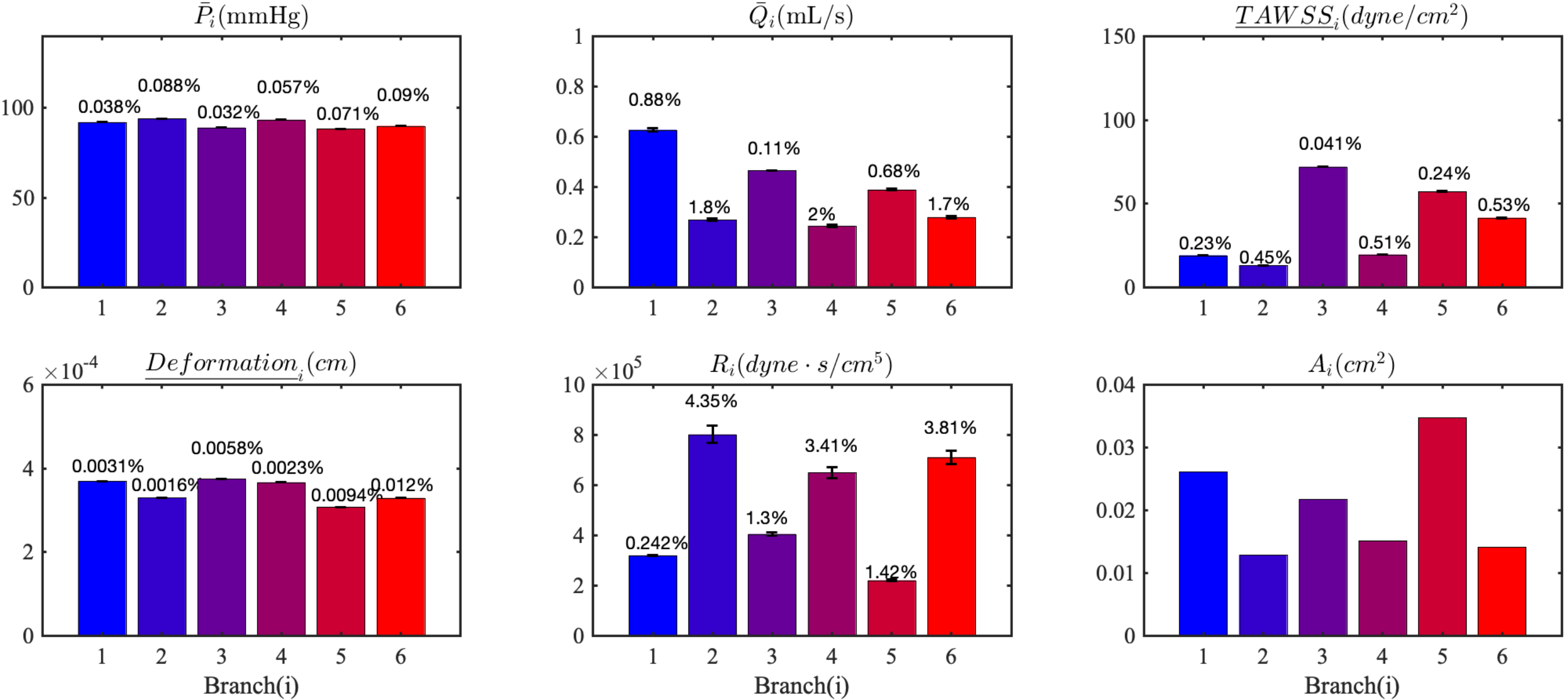}
\caption{Mean and standard deviations of LCA QoIs resulting from uncertainty in the morphometry law. The error bars represent two standard deviations. The coefficients of variation are reported at the top of each bar. $A_i$ is the cross-sectional area and $R_i$ is the resistance imposed at each outlet.}\label{fig:15}
\end{figure}

% ============================================================
\subsection{LCA: Effects of uncertainty in the material property, E$_s$}\label{sec:uqMat}
% ============================================================

\noindent We propagated the uncertainty in a spatially uniform, Gaussianly distributed elastic modulus, i.e., $\text{E}_s(\omega)\sim \mathcal{N}(1.48, 0.28^2)$, through the structural mechanics in our coronary vessel submodel. 
The results showed a $\overline{c}_v$ equal to 17\% in the wall deformation, while all other hemodynamic QoIs see limited changes, i.e., no more than 1\% (figure~\ref{fig:17}).
Note that, in our simulations, the magnitude of the wall deformation is small, only up to about one percent of the vessel radius, therefore not enough to affect flow and pressure (i.e., the hemodynamics).
The coefficient of variation in the wall deformation directly corresponds to $\overline{c}_v=$ 17\% in the imposed material property uncertainty, due to the direct proportionality between $p$ and $\delta_{r}$, discussed in the previous sections. 
This result suggests a loose coupling between hemodynamics and solid mechanics, where uncertainties tend to affect one of the two physics, but not both, as previously observed in~\cite{Tran2019}.

% ===================================================================
\subsection{LCA: Effects of uncertainty in P$_{\text{in}}$, P$_{\text{im,t}}$, $m$, E$_s$}\label{sec:uqAll}
% ===================================================================

\begin{figure}[ht]
\vspace{-6pt}
\centering
\includegraphics[width=0.8\textwidth, keepaspectratio]{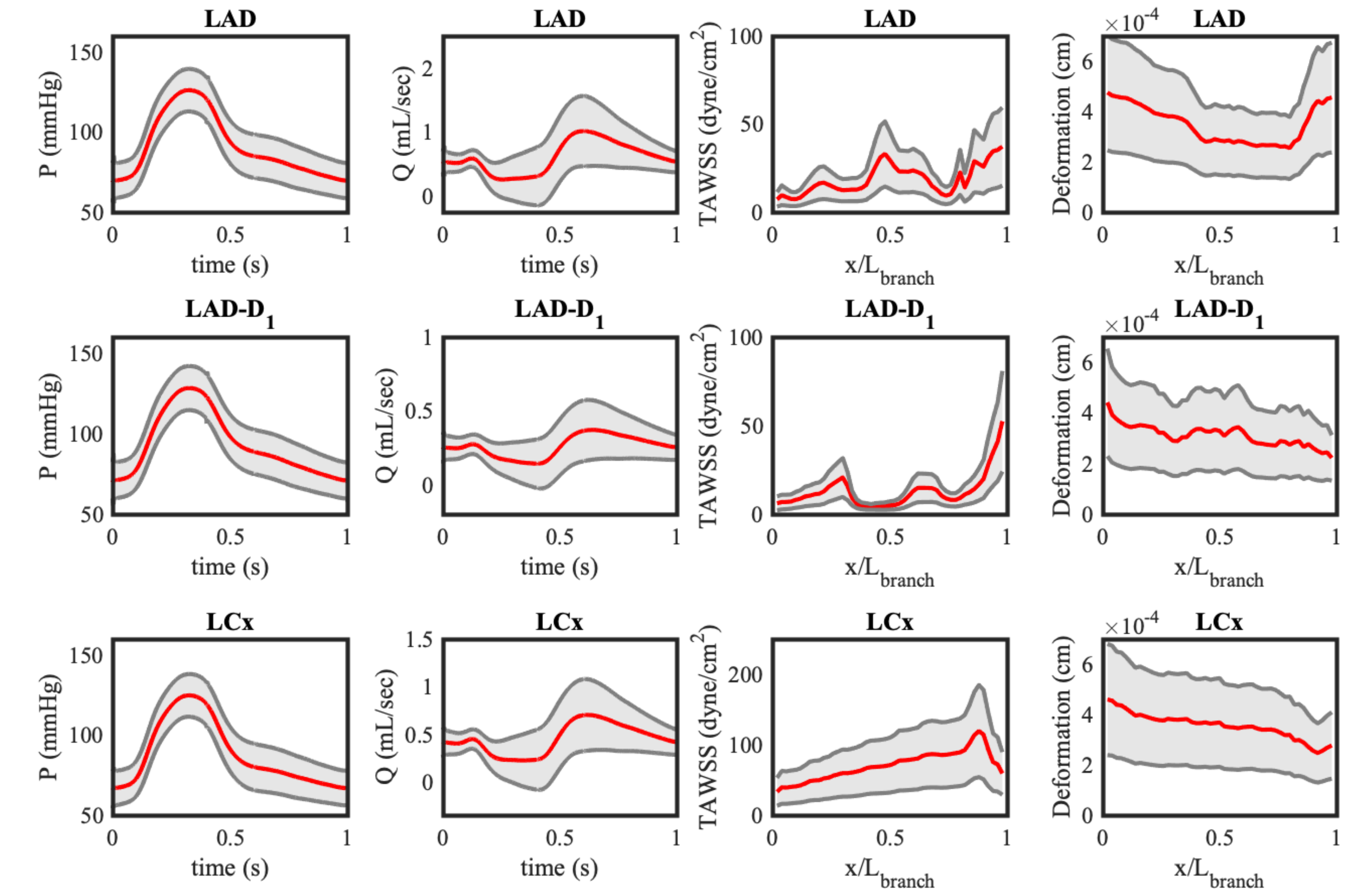}
\includegraphics[width=0.85\textwidth, keepaspectratio]{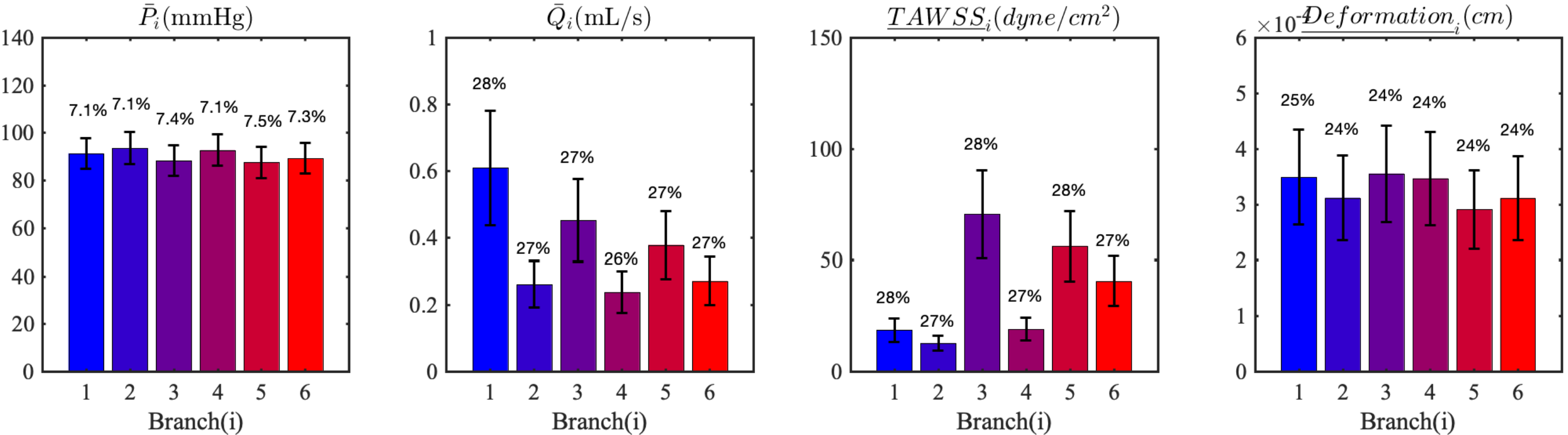}
\caption{Statistics of the QoIs in six branches resulting from all perturbed inputs. Ensemble averaged quantities are plotted with red lines, and 95 percent confidence intervals are plotted in gray. (Bar plots) Mean and standard deviations of LCA QoI resulting from uncertainty in all inputs. The error bars represent two standard deviation. The coefficients of variation are reported at the top of each bar.}\label{fig:17}
\end{figure}

\noindent 
Lastly, we simultaenously perturbed all input parameters including P$_{\text{in}}$, P$_{\text{im,t}}$, $m$, E$_s$. Figure~\ref{fig:17} shows the combined effects of all input parameter perturbations on the resulting variabilities in QoIs. The pressure uncertainty up to 7 percent $\bar{c}_v$ is mainly caused by the perturbation of P$_{\text{in}}$ since other input parameters do not affect the P$_{\text{in}}$ significantly, as shown in Section \ref{sec:uqIntra} - \ref{sec:uqMat}. The flow and TWASS variabilities, $\bar{c}_v\approx 27\%$,  are mainly governed by the perturbation of P$_\text{im,t}$ and the effect of perturbation of P$_\text{in}$ is not cumulatively added. However, the variabilty in deformation, $\bar{c}_v\approx 24\%$, appears to be a cumulative sum of resulting variabilities in deformation from the perturbations in P$_\text{in}$ and E$_s$, which was $\approx 7\%$ and $\approx 17\%$, respectively.

%
% ==========================================================
\subsection{Performance of UQ methodologies}\label{sec:uqPerformance}
% ==========================================================
%
\begin{figure}[ht]
\vspace{-6pt}
\centering
\includegraphics[width=1.0\textwidth, keepaspectratio]{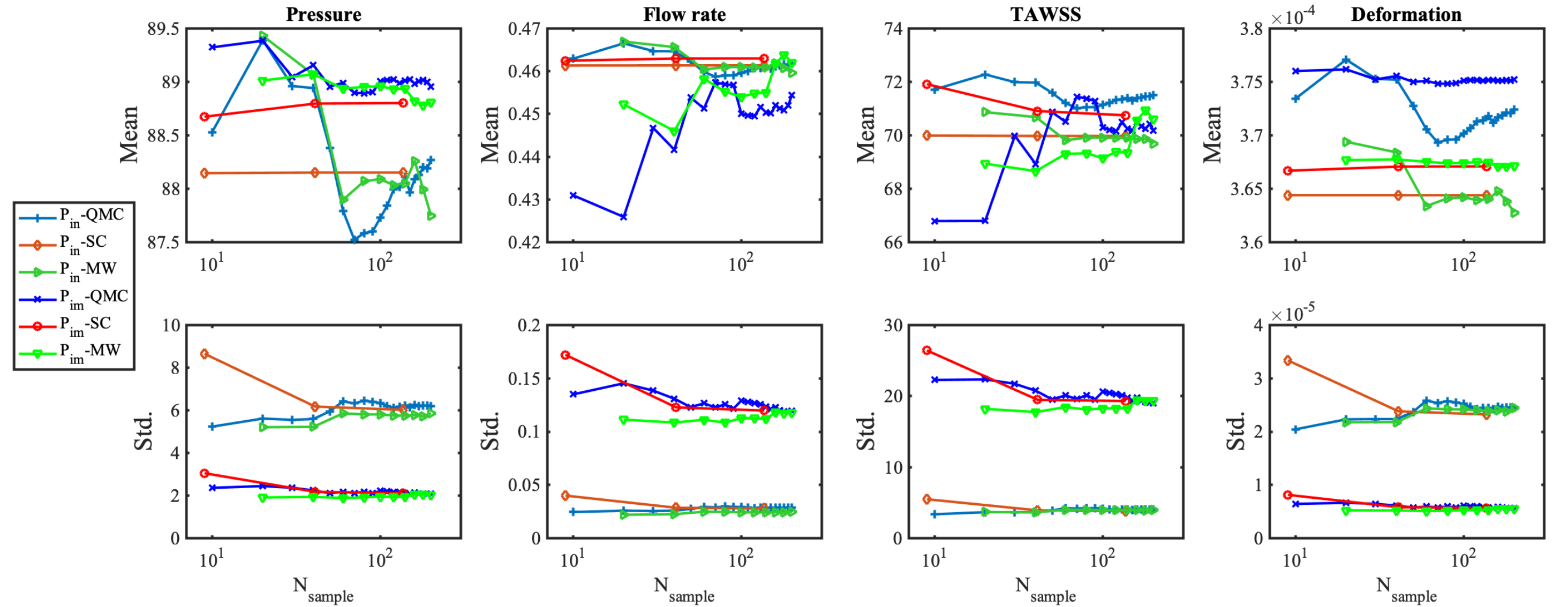}
\mylab{-72mm}{66mm}{(a)}%
\mylab{-34mm}{66mm}{(b)}%
\mylab{6mm}{66mm}{(c)}%
\mylab{44mm}{66mm}{(d)}%
\mylab{-68mm}{35mm}{(e)}%
\mylab{-32mm}{35mm}{(f)}%
\mylab{5mm}{35mm}{(g)}%
\mylab{44mm}{35mm}{(h)}%
\caption{Convergence of mean quantities of interest (a-d) and standard deviations (e-h) resulting from perturbing the inlet pressure and the intramyocardial pressure.}\label{fig:18}
\end{figure}

\begin{figure}[ht]
\vspace{-6pt}
\centering
\includegraphics[width=1.0\textwidth, keepaspectratio]{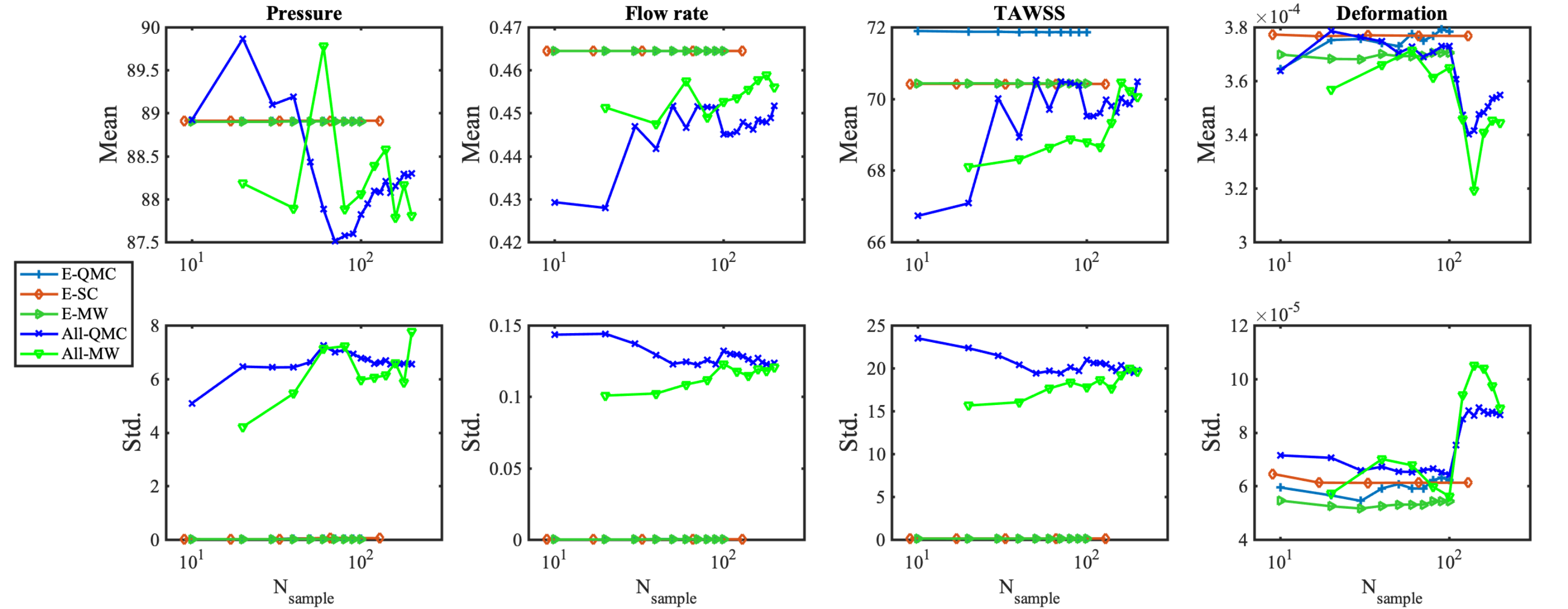}
\mylab{-72mm}{67mm}{(a)}%
\mylab{-35mm}{67mm}{(b)}%
\mylab{5mm}{67mm}{(c)}%
\mylab{43mm}{67mm}{(d)}%
\mylab{-68mm}{36mm}{(e)}%
\mylab{-32mm}{36mm}{(f)}%
\mylab{5mm}{36mm}{(g)}%
\mylab{44mm}{36mm}{(h)}%
\caption{Convergence of mean quantities of interest (a-d) and standard deviations (e-h) resulting from perturbing the Young's modulus, and all input parameters simultaneously.}\label{fig:19}
\end{figure}

\noindent As previously discussed in Section~\ref{sec:UQdetails}, we compare the performance of several UQ propagation approaches applied to our coronary model problem. 
Convergence in the QoI means and standard deviations are plotted in Figure~\ref{fig:18} and Figure~\ref{fig:19}, versus the number of associated model evaluations.

For uncertainty in the inlet pressure P$_\text{in}$, we observe that QMC converges slowly to the solution, and moment estimates for TAWSS and displacements are off the converged values provided by other methods.
With only 9 samples, the SC produces accurate estimates of the mean QoIs, but higher order integration with 41 samples is necessary to capture the standard deviation with sufficient accuracy.
MW on QMC grid provided a better estimate of the statistics overall, even for TAWSS and wall deformation.

% Results from the intramyocardial pressure uncertainty
For an uncertain P$_\text{im,t}$, we observed fast convergence of SC after level 2, consistent with the case above with random P$_\text{in}$, while MW estimates confirmed to be superior to QMC, using the same underlying samples. 
With the uncertain E$_\text{s}$, all methods converge quickly after 10 samples. QMC showed consistent errors when estimating mean of TAWSS, while MW corrects the mean estimates. MW showed lower mean and standard deviations for the displacement measurement. 
When all parameters are simultaneously perturbed, MW has less error at relatively low number of samples below 100 than QMC.
Finally, we find that, for the analyzed model problem, an accurate measure of the first two moments of the QoIs requires approximately at least $50$ samples.

% ================================
\section{Discussion}\label{sec:Discussion}
% ================================

\noindent Recent advances in model-based diagnostics of coronary artery disease have demonstrated how models can be successfully integrated in clinical routines, but have raised questions related to their robustness, in light of possible uncertainty or ignorance associated with their input parameters. 
In this study, we focus on four clinically relevant parameters, coronary artery pressure, intramyocardial pressures, wall stiffness, and the morphometry exponent, which need to be invariably specified for coronary models.
Uncertainty is injected in these parameters through stochastic processes in time, represented by a finite collection of independent random variables following Karhunen-Lo\`eve expansion. 

In this study, uncertainty in the intra-coronary pressure waveform was directly estimated using repeated measurements from cardiac catheterization in six patients. The pressure waveform data were characterized by a different mean but a consistent 5\% to 7\% coefficient of variation at all times.
We also considered uncertainty in another important boundary condition parameter, the intramyocardial pressure time-derivative, assuming a 10\% variability with respect to the maximum absolute value of this derivative during the cardiac cycle.
The morphometry exponent and the vessel elastic modulus are modeled as random variables whose moments are inferred from literature data on human coronary physiology.

In our forward propagation study, we use a multi-scale cardiovascular model with deformable walls and arbitrary Lagrangian Eulerian (ALE) fluid-structure interaction with conforming fluid and wall mesh interfaces.
A sub-model representing the left-coronary artery and its branches was extracted from a full aorto-coronary model, thus providing significant computational cost savings. Despite common perceptions that UQ is too expensive for large-scale problems, we demonstrated application of several UQ propagation strategies to a patient-specific ALE model, facilitated by the use of a sub-model to reduce cost.
We ran 1203 simulations in parallel, and identified four quantities of interest including branch pressure, flow rate, time-averaged wall shear stress, and wall deformation. 

Simulation results reveal that 7\% $\overline{c}_v$ in the inlet pressure produces 7\% $\overline{c}_v$ in pressure and wall deformation at all branches, while flow rate and wall shear stress outputs are found to be associated to variabilities (i.e., $\overline{c}_v$) of approximately 5\%.
Moreover, 10\% $\overline{c}_v$ in the intramyocardial pressure led instead to 27\% $\overline{c}_v$ in the flow rate and wall shear stress, while pressure and deformation variability remains limited to 3\% $\overline{c}_v$ in all branches.
Conversely, an uncertain morphometry exponent $m$ uniformly distributed between 2.4 to 2.8, seems not to affect simulation results. 
A Gaussian uncertainty with 17\% $\overline{c}_v$ in the Young's modulus of the vessel wall material seems to only affect the wall deformation QoI with equal 17\% variability, leaving other QoIs unperturbed.

The significant variability of TAWSS (27\% $\overline{c}_v$) due to small perturbations of the intramyocardial pressure (10\% $\overline{c}_v$) suggests that a reduction in the intramyocardial pressure uncertainty is important to provide accurate estimates of TAWSS in coronary artery simulations.
TAWSS has been identified as a major factor implicated in intimal thickening, and the progression of atherosclerosis and its rupture in coronary arteries \cite{Friedman1987, Gibson1993, Habib2011}. Recent studies \cite{Timmins2015, Timmins2016, Timmins2017} have demonstrated that low TAWSS computed from CFD is well correlated with coronary artery plaque progression. In future applications of CFD for prediction of atherosclerotic plaque rupture, ignoring variability of TAWSS may lead to inaccurate risk stratification of atherosclerosis. 
Also, the variability of pressure in coronary artery due to the corresponding inlet pressure uncertainty may lead to variability of FFR$_\text{CT}$ in coronary artery models.

% Compared the performance of UP approaches
A main purpose of this study is to compare the performance of several approaches for uncertainty propagation, Monte Carlo and Quasi Monte Carlo sampling, Stochastic Collocation, and multi-resolution stochastic expansion.
In this context, the performance of each approach was tested on three analytic models and one non-linear equation system. 
In the test problems, SC showed fast convergence on smooth response surfaces, but failed to provide accurate estimates of the statistical moments for response surfaces containing sharp gradients or discontinuities. 
Also QMC produced better convergence rates than MC, and MW produced convergence rates comparable to QMC. In high-dimensional problems, MW showed superior performance in estimating mean, and showed slower convergence in estimating standard deviations than QMC. The MW showed but showed the best performance on discontinuous response surfaces. 
Finally, we investigated and compared the convergence of QMC, SC and MW on a patient-specific coronary model. We find that SC requires a sufficient level refinement to get correct standard deviation estimates, and QMC shows slow convergence and incorrect estimates for TAWSS and wall deformations. MW shows superior performance and accuracy over the QMC method, and and better performance than SC especially with the small number of samples below 50. All in all, the MW method achieved the best performance of the methods compared, achieving fast convergence without sacrificing accuracy.  

%All different methods converged in the low order statistics. 
This study demonstrates the effects of uncertainty in the selection of four very important parameters affecting coronary flows, and takes an important step in comparing competing methods for uncertainty propagation on a realistic 3D problem.  
However, we recognize several limitations. 
% Not all sources of uncertainty
This study did not investigate all the possible sources of uncertainty in coronary models and additional sources should be added in future studies. 
% Geometrical uncertainty
One of the most important sources, for example, relates to uncertainty in the lumen diameter due to limitations in the resolution of the selected imaging modality and operator-dependent segmentation. 
The effects of this type of uncertainty are particularly relevant on stenotic coronary anatomies, which justifies our choice of analyzing a model without significant lumen reductions, and to withhold consideration of output quantities such as FFR until future work.
% Independent parameters
All uncertain parameters were considered independent, and interactions among them have been neglected. A future study will quantify the correlations among uncertain inputs and their interactions.
% Simplified isotropic material properties
An isotropic and homogeneous elastic vessel wall was assumed instead of more accurate descriptions, for example characterized by three-layer models consisting of intima, media, and adventitia with distinct material properties~\cite{Holzapfel2005}, or using more sophisticated hyperelastic constitutive models~\cite{Zhou1997, Kural2012}. 
Additionally, the distribution of the random inputs were assumed in this study rather than inferred from available clinical data, as discussed in some of the studies in the literature~\cite{Tran2019}. 
We selected the wall displacement as the mechanical output of interest and neglected both the pre-stress in the vessel walls and the extra stiffness (assumed negligible) associated with it.
% Growth and remodeling 
Finally, tissue growth and remodeling due to biological adaptation was not considered, though uncertainty in these models has been considered in prior studies. 

% Future work
Future work will also focus on systematic use of new estimators, such as multi-fidelity control variate estimators that show significant promise in computational cost saving for computationally expensive models \cite{Geraci2015}. The multi-fidelity approach has been recently applied to cardiovascular simulations \cite{Fleeter2017, Seo2020} to achieve significant improvements of accuracy and variance reduction, leveraging the low computational cost of reduced-order models (e.g. one-dimensional wave propagation model or zero-dimensional lumped parameter model), with a fraction of the three-dimensional model cost.
% ======================
\section{Acknowledgements}
% ======================

\noindent This work was supported by NIH grant (NIH NIBIB R01-EB018302), NSF SSI grants 1663671 and 1339824, and NSF CDSE CBET 1508794. This work used the Extreme Science and Engineering Discovery Environment (XSEDE)\cite{XSEDE}, which is supported by National Science Foundation grant number ACI-1548562.
We thank Mahidhar Tatineni for assisting on building Trilinos on Comet cluster, which was made possible through the XSEDE Extended Collaborative Support Service (ECSS) program \cite{EECS}. 
The authors also thank Professor Tran for fruitful discussions that helped in the preparation of this paper. We also acknowledge support from the open source SimVascular project at www.simvascular.org.
\appendix

\bibliographystyle{abbrv} 
\bibliography{UQpapers}  

\end{document}